\documentclass[pre,eqsecnum,preprint,showpacs,preprintnumbers,amsmath,amssymb,amsfonts]{revtex4}

\input epsf

\usepackage{graphicx}
\usepackage{bm}

\newcommand{\beq}{\begin{equation}}
\newcommand{\eeq}{\end{equation}}
\newcommand{\beqs}{\begin{eqnarray}}
\newcommand{\eeqs}{\end{eqnarray}}

\newtheorem{cor}{Corollary}[section]
\newtheorem{lemma}{Lemma}[section]
\newtheorem{defi}{Definition}[section]
\newtheorem{conj}{Conjecture}[section]
\newtheorem{propo}{Proposition}[section]

\begin{document}

\title{Dimer-monomer model on the Sierpinski gasket}

\author{Shu-Chiuan Chang}
\email{scchang@mail.ncku.edu.tw}
\affiliation{Department of Physics \\
National Cheng Kung University \\
Tainan 70101, Taiwan}

\author{Lung-Chi Chen}
\email{lcchen@math.fju.edu.tw}
\affiliation{Department of Mathematics \\
Fu Jen Catholic University \\
Taipei 24205, Taiwan}

\date{\today}

\begin{abstract}

We present the numbers of dimer-monomers on the Sierpinski gasket $SG_d(n)$ at stage $n$ with dimension $d$ equal to two, three and four, and determine the asymptotic behaviors. The corresponding results on the generalized Sierpinski gasket $SG_{d,b}(n)$ with $d=2$ and $b=3,4$ are obtained. 


\end{abstract}

\maketitle

\section{Introduction}
\label{sectionI}

The enumeration of the number of dimer-monomers $N_{DM}(G)$ on a graph $G$ is a classical model \cite{gaunt69,Heilmann70,Heilmann72}. In the model, each diatomic molecule is regarded as a dimer which occupies two adjacent sites of the graph. The sites that are not covered by any dimers are considered as occupied by monomers. Although the close-packed dimer problem on planar lattices has been expressed in closed-form almost half a century ago \cite{kasteleyn61,temperley61,fisher61}, the general dimer-monomer problem was shown to be computationally intractable \cite{jerrum}.
Some recent studies on the enumeration of close-packed dimer, single-monomer and general dimer-monomer problems on regular lattices were carried out in Refs. \cite{lu99,tzeng03,izmailian03,izmailian05,yan05,yan06,kong06,izmailian06,wu06,kong06n,kong06nn}. It is of interest to consider dimer-monomers on self-similar fractal lattices which have scaling invariance rather than translational invariance. Fractals are geometric structures of non-integer Hausdorff dimension realized by repeated construction of an elementary shape on progressively smaller length scales \cite{mandelbrot,Falconer}. A well-known example of fractal is the Sierpinski gasket which has been extensively studied in several contexts \cite{Gefen80,Gefen81,Rammal,Alexander,Domany,Gefen8384,Guyer,Kusuoka,Dhar97,Daerden,Dhar05}. We shall derive the recursion relations for the numbers of dimer-monomers on the Sierpinski gasket with dimension equal to two, three and four, and determine the asymptotic growth constants. We shall also consider the number of dimer-monomers on the generalized Sierpinski gasket with dimension equal to two.

\section{Preliminaries}
\label{sectionII}

We first recall some relevant definitions in this section. A connected graph (without loops) $G=(V,E)$ is defined by its vertex (site) and edge (bond) sets $V$ and $E$ \cite{bbook,fh}.  Let $v(G)=|V|$ be the number of vertices and $e(G)=|E|$ the number of edges in $G$.  The degree or coordination number $k_i$ of a vertex $v_i \in V$ is the number of edges attached to it.  A $k$-regular graph is a graph with the property that each of its vertices has the same degree $k$. In general, one can associate monomer and dimer weights to each monomer and dimer connecting adjacent vertices (see, for example \cite{wu06}). For simplicity, all such weights are set to one throughout this paper. 

When the number of dimer-monomers $N_{DM}(G)$ grows exponentially with $v(G)$ as $v(G) \to \infty$, there exists a constant $z_G$ describing this exponential growth:
\beq
z_G = \lim_{v(G) \to \infty} \frac{\ln N_{DM}(G)}{v(G)} \ ,
\label{zdef}
\eeq
where $G$, when used as a subscript in this manner, implicitly refers to
the thermodynamic limit.

\begin{figure}[htbp]
\unitlength 0.9mm \hspace*{3mm}
\begin{picture}(108,40)
\put(0,0){\line(1,0){6}}
\put(0,0){\line(3,5){3}}
\put(6,0){\line(-3,5){3}}
\put(3,-4){\makebox(0,0){$SG_2(0)$}}
\put(12,0){\line(1,0){12}}
\put(12,0){\line(3,5){6}}
\put(24,0){\line(-3,5){6}}
\put(15,5){\line(1,0){6}}
\put(18,0){\line(3,5){3}}
\put(18,0){\line(-3,5){3}}
\put(18,-4){\makebox(0,0){$SG_2(1)$}}
\put(30,0){\line(1,0){24}}
\put(30,0){\line(3,5){12}}
\put(54,0){\line(-3,5){12}}
\put(36,10){\line(1,0){12}}
\put(42,0){\line(3,5){6}}
\put(42,0){\line(-3,5){6}}
\multiput(33,5)(12,0){2}{\line(1,0){6}}
\multiput(36,0)(12,0){2}{\line(3,5){3}}
\multiput(36,0)(12,0){2}{\line(-3,5){3}}
\put(39,15){\line(1,0){6}}
\put(42,10){\line(3,5){3}}
\put(42,10){\line(-3,5){3}}
\put(42,-4){\makebox(0,0){$SG_2(2)$}}
\put(60,0){\line(1,0){48}}
\put(72,20){\line(1,0){24}}
\put(60,0){\line(3,5){24}}
\put(84,0){\line(3,5){12}}
\put(84,0){\line(-3,5){12}}
\put(108,0){\line(-3,5){24}}
\put(66,10){\line(1,0){12}}
\put(90,10){\line(1,0){12}}
\put(78,30){\line(1,0){12}}
\put(72,0){\line(3,5){6}}
\put(96,0){\line(3,5){6}}
\put(84,20){\line(3,5){6}}
\put(72,0){\line(-3,5){6}}
\put(96,0){\line(-3,5){6}}
\put(84,20){\line(-3,5){6}}
\multiput(63,5)(12,0){4}{\line(1,0){6}}
\multiput(66,0)(12,0){4}{\line(3,5){3}}
\multiput(66,0)(12,0){4}{\line(-3,5){3}}
\multiput(69,15)(24,0){2}{\line(1,0){6}}
\multiput(72,10)(24,0){2}{\line(3,5){3}}
\multiput(72,10)(24,0){2}{\line(-3,5){3}}
\multiput(75,25)(12,0){2}{\line(1,0){6}}
\multiput(78,20)(12,0){2}{\line(3,5){3}}
\multiput(78,20)(12,0){2}{\line(-3,5){3}}
\put(81,35){\line(1,0){6}}
\put(84,30){\line(3,5){3}}
\put(84,30){\line(-3,5){3}}
\put(84,-4){\makebox(0,0){$SG_2(3)$}}
\end{picture}

\vspace*{5mm}
\caption{\footnotesize{The first four stages $n=0,1,2,3$ of the two-dimensional Sierpinski gasket $SG_2(n)$.}} 
\label{sgfig}
\end{figure}
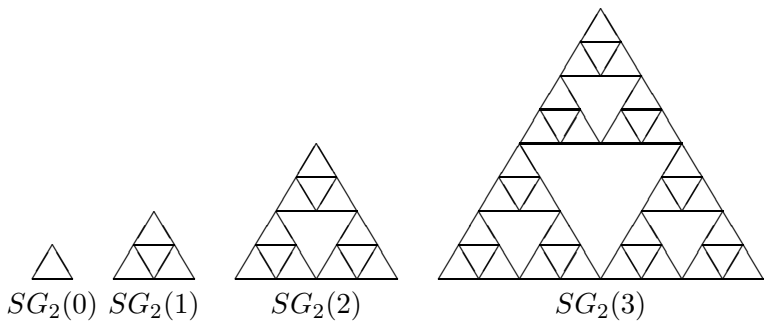

The construction of the two-dimensional Sierpinski gasket $SG_2(n)$ at stage $n$ is shown in Fig. \ref{sgfig}. At stage $n=0$, it is an equilateral triangle; while stage $n+1$ is obtained by the juxtaposition of three $n$-stage structures. In general, the Sierpinski gaskets $SG_d$ can be built in any Euclidean dimension $d$ with fractal dimensionality $D=\ln(d+1)/\ln2$ \cite{Gefen81}. For the Sierpinski gasket $SG_d(n)$, the numbers of edges and vertices are given by 
\beq
e(SG_d(n)) = {d+1 \choose 2} (d+1)^n = \frac{d}{2} (d+1)^{n+1} \ ,
\label{e}
\eeq
\beq
v(SG_d(n)) = \frac{d+1}{2} [(d+1)^n+1] \ .
\label{v}
\eeq
Except the $(d+1)$ outmost vertices which have degree $d$, all other vertices of $SG_d(n)$ have degree $2d$. In the large $n$ limit, $SG_d$ is $2d$-regular. 

\bigskip

\bigskip

\begin{figure}[htbp]
\unitlength 0.9mm \hspace*{3mm}
\begin{picture}(108,45)
\put(0,0){\line(1,0){18}}
\put(3,5){\line(1,0){12}}
\put(6,10){\line(1,0){6}}
\put(0,0){\line(3,5){9}}
\put(6,0){\line(3,5){6}}
\put(12,0){\line(3,5){3}}
\put(18,0){\line(-3,5){9}}
\put(12,0){\line(-3,5){6}}
\put(6,0){\line(-3,5){3}}
\put(9,-4){\makebox(0,0){$SG_{2,3}(1)$}}
\put(24,0){\line(1,0){54}}
\put(33,15){\line(1,0){36}}
\put(42,30){\line(1,0){18}}
\put(24,0){\line(3,5){27}}
\put(42,0){\line(3,5){18}}
\put(60,0){\line(3,5){9}}
\put(78,0){\line(-3,5){27}}
\put(60,0){\line(-3,5){18}}
\put(42,0){\line(-3,5){9}}
\multiput(27,5)(18,0){3}{\line(1,0){12}}
\multiput(30,10)(18,0){3}{\line(1,0){6}}
\multiput(30,0)(18,0){3}{\line(3,5){6}}
\multiput(36,0)(18,0){3}{\line(3,5){3}}
\multiput(36,0)(18,0){3}{\line(-3,5){6}}
\multiput(30,0)(18,0){3}{\line(-3,5){3}}
\multiput(36,20)(18,0){2}{\line(1,0){12}}
\multiput(39,25)(18,0){2}{\line(1,0){6}}
\multiput(39,15)(18,0){2}{\line(3,5){6}}
\multiput(45,15)(18,0){2}{\line(3,5){3}}
\multiput(45,15)(18,0){2}{\line(-3,5){6}}
\multiput(39,15)(18,0){2}{\line(-3,5){3}}
\put(45,35){\line(1,0){12}}
\put(48,40){\line(1,0){6}}
\put(48,30){\line(3,5){6}}
\put(54,30){\line(3,5){3}}
\put(54,30){\line(-3,5){6}}
\put(48,30){\line(-3,5){3}}
\put(48,-4){\makebox(0,0){$SG_{2,3}(2)$}}
\put(84,0){\line(1,0){24}}
\put(87,5){\line(1,0){18}}
\put(90,10){\line(1,0){12}}
\put(93,15){\line(1,0){6}}
\put(84,0){\line(3,5){12}}
\put(90,0){\line(3,5){9}}
\put(96,0){\line(3,5){6}}
\put(102,0){\line(3,5){3}}
\put(108,0){\line(-3,5){12}}
\put(102,0){\line(-3,5){9}}
\put(96,0){\line(-3,5){6}}
\put(90,0){\line(-3,5){3}}
\put(96,-4){\makebox(0,0){$SG_{2,4}(1)$}}
\end{picture}

\vspace*{5mm}
\caption{\footnotesize{The generalized two-dimensional Sierpinski gasket $SG_{2,b}(n)$ with $b=3$ at stage $n=1, 2$ and $b=4$ at stage $n=1$.}} 
\label{sgbfig}
\end{figure}
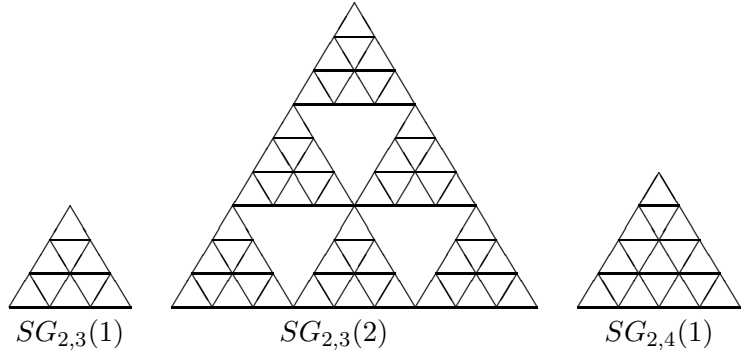

The Sierpinski gasket can be generalized, denoted as $SG_{d,b}(n)$, by introducing the side length $b$ which is an integer larger or equal to two \cite{Hilfer}. The generalized Sierpinski gasket at stage $n+1$ is constructed with $b$ layers of stage $n$ hypertetrahedrons. The two-dimensional $SG_{2,b}(n)$ with $b=3$ at stage $n=1, 2$ and $b=4$ at stage $n=1$ are illustrated in Fig. \ref{sgbfig}. The ordinary Sierpinski gasket $SG_d(n)$ corresponds to the $b=2$ case, where the index $b$ is neglected for simplicity. The Hausdorff dimension for $SG_{d,b}$ is given by $D=\ln {b+d-1 \choose d} / \ln b$ \cite{Hilfer}. Notice that $SG_{d,b}$ is not $k$-regular even in the thermodynamic limit.

\bigskip

\bigskip

\section{The number of dimer-monomers on $SG_2(n)$}
\label{sectionIII}

In this section we derive the asymptotic growth constant for the number of dimer-monomers on the two-dimensional Sierpinski gasket $SG_2(n)$ in detail. Let us start with the definitions of the quantities to be used.

\bigskip

\begin{defi} \label{defisg2} Consider the generalized two-dimensional Sierpinski gasket $SG_{2,b}(n)$ at stage $n$. (a) Define $M_{2,b}(n) \equiv N_{DM}(SG_{2,b}(n))$ as the number of dimer-monomers. (b) Define $f_{2,b}(n)$ as the number of dimer-monomers such that the three outmost vertices are occupied by monomers. (c) Define $gr_{2,b}(n)$, $gl_{2,b}(n)$, $gt_{2,b}(n)$ as the numbers of dimer-monomers such that either rightmost, leftmost or topmost vertex, respectively, is occupied by a dimer and the other two outmost vertices are occupied by monomers. (d) Define $hr_{2,b}(n)$, $hl_{2,b}(n)$, $ht_{2,b}(n)$ as the numbers of dimer-monomers such that either rightmost, leftmost or topmost vertex, respectively, is occupied by a monomer and the other two outmost vertices are occupied by dimers. (e) Define $t_{2,b}(n)$ as the number of dimer-monomers such that all three outmost vertices are occupied by dimers. 
\end{defi}

\bigskip

\begin{figure}[htbp]
\unitlength 1.8mm 
\begin{picture}(54,5)
\put(3,1.7){\circle{2}}
\put(0,0){\line(1,0){6}}
\put(0,0){\line(3,5){3}}
\put(6,0){\line(-3,5){3}}
\put(3,-2){\makebox(0,0){$M_2(n)$}}
\put(12,0){\line(1,0){6}}
\put(12,0){\line(3,5){3}}
\put(18,0){\line(-3,5){3}}
\multiput(13,0.5)(4,0){2}{\circle{1}}
\put(15,4){\circle{1}}
\put(15,-2){\makebox(0,0){$f_2(n)$}}
\put(24,0){\line(1,0){6}}
\put(24,0){\line(3,5){3}}
\put(30,0){\line(-3,5){3}}
\put(25,0.5){\circle{1}}
\put(29,0.5){\circle*{1}}
\put(27,4){\circle{1}}
\put(27,-2){\makebox(0,0){$gr_2(n)$}}
\put(36,0){\line(1,0){6}}
\put(36,0){\line(3,5){3}}
\put(42,0){\line(-3,5){3}}
\put(37,0.5){\circle*{1}}
\put(41,0.5){\circle{1}}
\put(39,4){\circle{1}}
\put(39,-2){\makebox(0,0){$gl_2(n)$}}
\put(48,0){\line(1,0){6}}
\put(48,0){\line(3,5){3}}
\put(54,0){\line(-3,5){3}}
\multiput(49,0.5)(4,0){2}{\circle{1}}
\put(51,4){\circle*{1}}
\put(51,-2){\makebox(0,0){$gt_2(n)$}}
\end{picture}
\vspace*{1cm}

\begin{picture}(54,5)
\put(12,0){\line(1,0){6}}
\put(12,0){\line(3,5){3}}
\put(18,0){\line(-3,5){3}}
\put(13,0.5){\circle*{1}}
\put(17,0.5){\circle{1}}
\put(15,4){\circle*{1}}
\put(15,-2){\makebox(0,0){$hr_2(n)$}}
\put(24,0){\line(1,0){6}}
\put(24,0){\line(3,5){3}}
\put(30,0){\line(-3,5){3}}
\put(25,0.5){\circle{1}}
\put(29,0.5){\circle*{1}}
\put(27,4){\circle*{1}}
\put(27,-2){\makebox(0,0){$hl_2(n)$}}
\put(36,0){\line(1,0){6}}
\put(36,0){\line(3,5){3}}
\put(42,0){\line(-3,5){3}}
\put(37,0.5){\circle*{1}}
\put(41,0.5){\circle*{1}}
\put(39,4){\circle{1}}
\put(39,-2){\makebox(0,0){$ht_2(n)$}}
\put(48,0){\line(1,0){6}}
\put(48,0){\line(3,5){3}}
\put(54,0){\line(-3,5){3}}
\multiput(49,0.5)(4,0){2}{\circle*{1}}
\put(51,4){\circle*{1}}
\put(51,-2){\makebox(0,0){$t_2(n)$}}
\end{picture}

\vspace*{5mm}
\caption{\footnotesize{Illustration for the configurations $M_2(n)$, $f_2(n)$, $gr_2(n)$, $gl_2(n)$, $gt_2(n)$, $hr_2(n)$, $hl_2(n)$, $ht_2(n)$ and $t_2(n)$. Only the three outmost vertices are shown explicitly for $f_2(n)$, $g_2(n)$, $h_2(n)$ and $t_2(n)$, where each open circle is occupied by a monomer and each solid circle is occupied by a dimer.}} 
\label{fghtfig}
\end{figure}
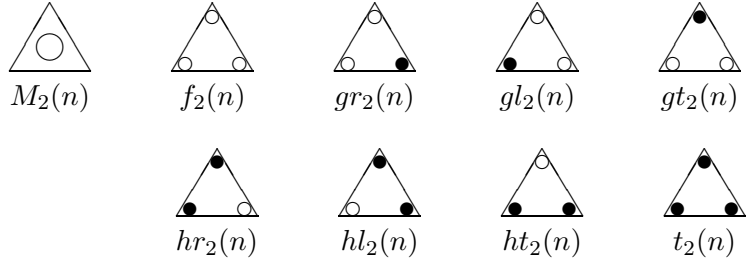

It is clear that the values $gr_{2,b}(n)$, $gl_{2,b}(n)$, $gt_{2,b}(n)$ are the same because of rotation symmetry, and we define $g_{2,b}(n) \equiv gr_{2,b}(n) = gl_{2,b}(n) = gt_{2,b}(n)$. Similarly, we define $h_{2,b}(n) \equiv hr_{2,b}(n) = hl_{2,b}(n) = ht_{2,b}(n)$. 
Since we only consider ordinary Sierpinski gasket in this section, we use the notations $M_2(n)$, $f_2(n)$, $g_2(n)$, $h_2(n)$, and $t_2(n)$ for simplicity. They are illustrated in Fig. \ref{fghtfig}, where only the outmost vertices are shown. It follows that
\beq
M_2(n) = f_2(n)+3g_2(n)+3h_2(n)+t_2(n) 
\label{Msg2}
\eeq
for non-negative integer $n$. The initial values at stage zero are $f_2(0)=1$, $g_2(0)=0$, $h_2(0)=1$, $t_2(0)=0$ and $M_2(0)=4$. The values at stage one are $f_2(1)=4$, $g_2(1)=4$, $h_2(1)=3$, $t_2(1)=2$ and $M_2(1)=27$. The purpose of this section is to obtain the asymptotic behavior of $M_2(n)$ as follows. The five quantities $M_2(n)$, $f_2(n)$, $g_2(n)$, $h_2(n)$ and $t_2(n)$ satisfy recursion relations. 

\bigskip

\bigskip

\begin{lemma} \label{lemmasg2r} For any non-negative integer $n$,
\beqs
M_2(n+1) & = & M_2^3(n) - 3M_2(n)[g_2(n)+2h_2(n)+t_2(n)]^2 \cr\cr & & + 3[h_2(n)+t_2(n)][g_2(n)+2h_2(n)+t_2(n)]^2 - [h_2(n)+t_2(n)]^3 \ , 
\label{Meq}
\eeqs
\beqs
f_2(n+1) & = & f_2^3(n) + 6f_2^2(n)g_2(n) + 3f_2^2(n)h_2(n) + 9f_2(n)g_2^2(n) + 2g_2^3(n) \cr\cr & & + 6f_2(n)g_2(n)h_2(n) \ , 
\label{feq}
\eeqs
\beqs
g_2(n+1) & = & f_2^2(n)g_2(n) + 2f_2^2(n)h_2(n) + 4f_2(n)g_2^2(n) + f_2^2(n)t_2(n) + 8f_2(n)g_2(n)h_2(n) \cr\cr & & + 3g_2^3(n) + 2f_2(n)g_2(n)t_2(n) + 2f_2(n)h_2^2(n) + 4g_2^2(n)h_2(n) \ , 
\label{geq}
\eeqs
\beqs
h_2(n+1) & = & f_2(n)g_2^2(n) + 4f_2(n)g_2(n)h_2(n) + 2g_2^3(n) + 2f_2(n)g_2(n)t_2(n) + 7g_2^2(n)h_2(n) \cr\cr & & + 3f_2(n)h_2^2(n) + 2f_2(n)h_2(n)t_2(n) + 2g_2^2(n)t_2(n) + 4g_2(n)h_2^2(n) \ ,
\label{heq}
\eeqs
\beqs
t_2(n+1) & = & g_2^3(n) + 6g_2^2(n)h_2(n) + 3g_2^2(n)t_2(n) + 9g_2(n)h_2^2(n) + 2h_2^3(n) \cr\cr & & + 6g_2(n)h_2(n)t_2(n) \ .
\label{teq}
\eeqs
\end{lemma}

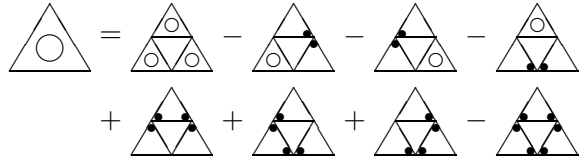
\begin{figure}[htbp]
\unitlength 0.9mm 
\begin{picture}(84,12)
\put(0,0){\line(1,0){12}}
\put(0,0){\line(3,5){6}}
\put(12,0){\line(-3,5){6}}
\put(6,3.4){\circle{4}}
\put(15,5){\makebox(0,0){$=$}}
\put(18,0){\line(1,0){12}}
\put(18,0){\line(3,5){6}}
\put(24,0){\line(3,5){3}}
\put(24,0){\line(-3,5){3}}
\put(21,5){\line(1,0){6}}
\put(30,0){\line(-3,5){6}}
\put(21,1.7){\circle{2}}
\put(27,1.7){\circle{2}}
\put(24,6.7){\circle{2}}
\put(33,5){\makebox(0,0){$-$}}
\put(36,0){\line(3,5){6}}
\put(36,0){\line(1,0){12}}
\put(42,0){\line(3,5){3}}
\put(42,0){\line(-3,5){3}}
\put(39,5){\line(1,0){6}}
\put(48,0){\line(-3,5){6}}
\put(39,1.7){\circle{2}}
\put(45,4){\circle*{1}}
\put(44,5.5){\circle*{1}}
\put(51,5){\makebox(0,0){$-$}}
\put(54,0){\line(3,5){6}}
\put(54,0){\line(1,0){12}}
\put(60,0){\line(3,5){3}}
\put(60,0){\line(-3,5){3}}
\put(57,5){\line(1,0){6}}
\put(66,0){\line(-3,5){6}}
\put(63,1.7){\circle{2}}
\put(57,4){\circle*{1}}
\put(58,5.5){\circle*{1}}
\put(69,5){\makebox(0,0){$-$}}
\put(72,0){\line(3,5){6}}
\put(72,0){\line(1,0){12}}
\put(78,0){\line(3,5){3}}
\put(78,0){\line(-3,5){3}}
\put(75,5){\line(1,0){6}}
\put(84,0){\line(-3,5){6}}
\put(78,6.7){\circle{2}}
\put(77,0.5){\circle*{1}}
\put(79,0.5){\circle*{1}}
\end{picture}

\unitlength 0.9mm 
\begin{picture}(84,12)
\put(15,5){\makebox(0,0){$+$}}
\put(18,0){\line(1,0){12}}
\put(18,0){\line(3,5){6}}
\put(24,0){\line(3,5){3}}
\put(24,0){\line(-3,5){3}}
\put(21,5){\line(1,0){6}}
\put(30,0){\line(-3,5){6}}
\put(27,4){\circle*{1}}
\put(26,5.5){\circle*{1}}
\put(21,4){\circle*{1}}
\put(22,5.5){\circle*{1}}
\put(33,5){\makebox(0,0){$+$}}
\put(36,0){\line(3,5){6}}
\put(36,0){\line(1,0){12}}
\put(42,0){\line(3,5){3}}
\put(42,0){\line(-3,5){3}}
\put(39,5){\line(1,0){6}}
\put(48,0){\line(-3,5){6}}
\put(39,4){\circle*{1}}
\put(40,5.5){\circle*{1}}
\put(41,0.5){\circle*{1}}
\put(43,0.5){\circle*{1}}
\put(51,5){\makebox(0,0){$+$}}
\put(54,0){\line(3,5){6}}
\put(54,0){\line(1,0){12}}
\put(60,0){\line(3,5){3}}
\put(60,0){\line(-3,5){3}}
\put(57,5){\line(1,0){6}}
\put(66,0){\line(-3,5){6}}
\put(63,4){\circle*{1}}
\put(62,5.5){\circle*{1}}
\put(59,0.5){\circle*{1}}
\put(61,0.5){\circle*{1}}
\put(69,5){\makebox(0,0){$-$}}
\put(72,0){\line(3,5){6}}
\put(72,0){\line(1,0){12}}
\put(78,0){\line(3,5){3}}
\put(78,0){\line(-3,5){3}}
\put(75,5){\line(1,0){6}}
\put(84,0){\line(-3,5){6}}
\put(77,0.5){\circle*{1}}
\put(79,0.5){\circle*{1}}
\put(81,4){\circle*{1}}
\put(80,5.5){\circle*{1}}
\put(75,4){\circle*{1}}
\put(76,5.5){\circle*{1}}
\end{picture}

\caption{\footnotesize{Illustration for the expression of  $M_2(n+1)$. Certain outmost vertices of $SG_2(n)$ are not shown means they can be occupied by either dimers or monomers.}} 
\label{Mfig}
\end{figure}

{\sl Proof} \quad 
The Sierpinski gaskets $SG_2(n+1)$ is composed of three $SG_2(n)$ with three pairs of vertices identified. For the number $M_2(n+1)$, the unallowable configurations are those with at least a pair of identified vertices originally occupied by dimers. Therefore, the three configuration with a pair of identified vertices occupied by dimers should be subtracted from all possible configurations $M_2^3(n)$. However, this procedure subtracts out also configurations with two pairs of identified vertices occupied by dimers that should be added back as illustrated in Fig. \ref{Mfig}. Finally, the configuration with three pairs of identified vertices occupied by dimers should be subtracted, and Eq. (\ref{Meq}) is verified.

\bigskip

\bigskip

\begin{figure}[htbp]
\unitlength 0.9mm 
\begin{picture}(136,12)
\put(0,0){\line(1,0){12}}
\put(0,0){\line(3,5){6}}
\put(12,0){\line(-3,5){6}}
\multiput(1,0.5)(10,0){2}{\circle{1}}
\put(6,9){\circle{1}}
\put(15,5){\makebox(0,0){$=$}}
\put(18,0){\line(1,0){12}}
\put(18,0){\line(3,5){6}}
\put(24,0){\line(3,5){3}}
\put(24,0){\line(-3,5){3}}
\put(21,5){\line(1,0){6}}
\put(30,0){\line(-3,5){6}}
\put(19,0.5){\circle{1}}
\put(21,4){\circle{1}}
\put(23,0.5){\circle{1}}
\put(22,5.5){\circle{1}}
\put(24,9){\circle{1}}
\put(26,5.5){\circle{1}}
\put(25,0.5){\circle{1}}
\put(27,4){\circle{1}}
\put(29,0.5){\circle{1}}
\put(36,5){\makebox(0,0){$+$}}
\put(39,0){\line(1,0){12}}
\put(39,0){\line(3,5){6}}
\put(45,0){\line(3,5){3}}
\put(45,0){\line(-3,5){3}}
\put(42,5){\line(1,0){6}}
\put(51,0){\line(-3,5){6}}
\put(40,0.5){\circle{1}}
\put(42,4){\circle{1}}
\put(44,0.5){\circle{1}}
\put(43,5.5){\circle{1}}
\put(45,9){\circle{1}}
\put(47,5.5){\circle*{1}}
\put(46,0.5){\circle{1}}
\put(48,4){\circle{1}}
\put(50,0.5){\circle{1}}
\put(52,5){\makebox(0,0){$\times 3$}}
\put(57,5){\makebox(0,0){$+$}}
\put(60,0){\line(1,0){12}}
\put(60,0){\line(3,5){6}}
\put(66,0){\line(3,5){3}}
\put(66,0){\line(-3,5){3}}
\put(63,5){\line(1,0){6}}
\put(72,0){\line(-3,5){6}}
\put(61,0.5){\circle{1}}
\put(63,4){\circle{1}}
\put(65,0.5){\circle{1}}
\put(64,5.5){\circle*{1}}
\put(66,9){\circle{1}}
\put(68,5.5){\circle{1}}
\put(67,0.5){\circle{1}}
\put(69,4){\circle{1}}
\put(71,0.5){\circle{1}}
\put(73,5){\makebox(0,0){$\times 3$}}
\put(78,5){\makebox(0,0){$+$}}
\put(81,0){\line(1,0){12}}
\put(81,0){\line(3,5){6}}
\put(87,0){\line(3,5){3}}
\put(87,0){\line(-3,5){3}}
\put(84,5){\line(1,0){6}}
\put(93,0){\line(-3,5){6}}
\put(82,0.5){\circle{1}}
\put(84,4){\circle{1}}
\put(86,0.5){\circle{1}}
\put(85,5.5){\circle*{1}}
\put(87,9){\circle{1}}
\put(89,5.5){\circle*{1}}
\put(88,0.5){\circle{1}}
\put(90,4){\circle{1}}
\put(92,0.5){\circle{1}}
\put(94,5){\makebox(0,0){$\times 3$}}
\put(99,5){\makebox(0,0){$+$}}
\put(102,0){\line(1,0){12}}
\put(102,0){\line(3,5){6}}
\put(108,0){\line(3,5){3}}
\put(108,0){\line(-3,5){3}}
\put(105,5){\line(1,0){6}}
\put(114,0){\line(-3,5){6}}
\put(103,0.5){\circle{1}}
\put(105,4){\circle{1}}
\put(107,0.5){\circle*{1}}
\put(106,5.5){\circle{1}}
\put(108,9){\circle{1}}
\put(110,5.5){\circle{1}}
\put(109,0.5){\circle{1}}
\put(111,4){\circle*{1}}
\put(113,0.5){\circle{1}}
\put(115,5){\makebox(0,0){$\times 3$}}
\put(120,5){\makebox(0,0){$+$}}
\put(123,0){\line(1,0){12}}
\put(123,0){\line(3,5){6}}
\put(129,0){\line(3,5){3}}
\put(129,0){\line(-3,5){3}}
\put(126,5){\line(1,0){6}}
\put(135,0){\line(-3,5){6}}
\put(124,0.5){\circle{1}}
\put(126,4){\circle*{1}}
\put(128,0.5){\circle{1}}
\put(127,5.5){\circle{1}}
\put(129,9){\circle{1}}
\put(131,5.5){\circle{1}}
\put(130,0.5){\circle*{1}}
\put(132,4){\circle{1}}
\put(134,0.5){\circle{1}}
\put(136,5){\makebox(0,0){$\times 3$}}
\end{picture}

\begin{picture}(136,12)
\put(15,5){\makebox(0,0){$+$}}
\put(18,0){\line(1,0){12}}
\put(18,0){\line(3,5){6}}
\put(24,0){\line(3,5){3}}
\put(24,0){\line(-3,5){3}}
\put(21,5){\line(1,0){6}}
\put(30,0){\line(-3,5){6}}
\put(19,0.5){\circle{1}}
\put(21,4){\circle*{1}}
\put(23,0.5){\circle{1}}
\put(22,5.5){\circle{1}}
\put(24,9){\circle{1}}
\put(26,5.5){\circle{1}}
\put(25,0.5){\circle{1}}
\put(27,4){\circle*{1}}
\put(29,0.5){\circle{1}}
\put(31,5){\makebox(0,0){$\times 3$}}
\put(36,5){\makebox(0,0){$+$}}
\put(39,0){\line(1,0){12}}
\put(39,0){\line(3,5){6}}
\put(45,0){\line(3,5){3}}
\put(45,0){\line(-3,5){3}}
\put(42,5){\line(1,0){6}}
\put(51,0){\line(-3,5){6}}
\put(40,0.5){\circle{1}}
\put(42,4){\circle{1}}
\put(44,0.5){\circle*{1}}
\put(43,5.5){\circle*{1}}
\put(45,9){\circle{1}}
\put(47,5.5){\circle{1}}
\put(46,0.5){\circle{1}}
\put(48,4){\circle*{1}}
\put(50,0.5){\circle{1}}
\put(57,5){\makebox(0,0){$+$}}
\put(60,0){\line(1,0){12}}
\put(60,0){\line(3,5){6}}
\put(66,0){\line(3,5){3}}
\put(66,0){\line(-3,5){3}}
\put(63,5){\line(1,0){6}}
\put(72,0){\line(-3,5){6}}
\put(61,0.5){\circle{1}}
\put(63,4){\circle*{1}}
\put(65,0.5){\circle{1}}
\put(64,5.5){\circle{1}}
\put(66,9){\circle{1}}
\put(68,5.5){\circle*{1}}
\put(67,0.5){\circle*{1}}
\put(69,4){\circle{1}}
\put(71,0.5){\circle{1}}
\put(78,5){\makebox(0,0){$+$}}
\put(81,0){\line(1,0){12}}
\put(81,0){\line(3,5){6}}
\put(87,0){\line(3,5){3}}
\put(87,0){\line(-3,5){3}}
\put(84,5){\line(1,0){6}}
\put(93,0){\line(-3,5){6}}
\put(82,0.5){\circle{1}}
\put(84,4){\circle*{1}}
\put(86,0.5){\circle{1}}
\put(85,5.5){\circle{1}}
\put(87,9){\circle{1}}
\put(89,5.5){\circle{1}}
\put(88,0.5){\circle*{1}}
\put(90,4){\circle*{1}}
\put(92,0.5){\circle{1}}
\put(94,5){\makebox(0,0){$\times 3$}}
\put(99,5){\makebox(0,0){$+$}}
\put(102,0){\line(1,0){12}}
\put(102,0){\line(3,5){6}}
\put(108,0){\line(3,5){3}}
\put(108,0){\line(-3,5){3}}
\put(105,5){\line(1,0){6}}
\put(114,0){\line(-3,5){6}}
\put(103,0.5){\circle{1}}
\put(105,4){\circle*{1}}
\put(107,0.5){\circle*{1}}
\put(106,5.5){\circle{1}}
\put(108,9){\circle{1}}
\put(110,5.5){\circle{1}}
\put(109,0.5){\circle{1}}
\put(111,4){\circle*{1}}
\put(113,0.5){\circle{1}}
\put(115,5){\makebox(0,0){$\times 3$}}
\end{picture}

\caption{\footnotesize{Illustration for the expression of  $f_2(n+1)$. The multiplication of three on the right-hand-side corresponds to the three possible orientations of $SG_2(n+1)$.}} 
\label{ffig}
\end{figure}
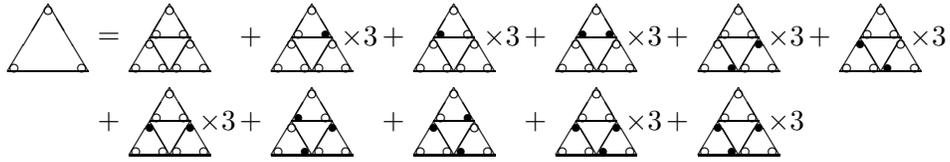

As illustrated in Fig. \ref{ffig}, the number $f_2(n+1)$ consists of (i) one configuration where all three of the $SG_2(n)$ are in the $f_2(n)$ status, (ii) six configurations where two of the $SG_2(n)$ are in the $f_2(n)$ status and the other one is in the $g_2(n)$ status, (iii) three configurations where two of the $SG_2(n)$ are in the $f_2(n)$ status and the other one is in the $h_2(n)$ status, (iv) nine configurations where one of the $SG_2(n)$ is in the $f_2(n)$ status and the other two are in the $g_2(n)$ status, (v) two configuration where all three of the $SG_2(n)$ are in the $g_2(n)$ status, (vi) six configurations where one of the $SG_2(n)$ is in the $f_2(n)$ status, another one is in the $g_2(n)$ status and the other one is in the $h_2(n)$ status. Therefore, we have
\beqs
f_2(n+1) & = & f_2^3(n) + 2f_2^2(n)[gr_2(n)+gl_2(n)+gt_2(n)] + f_2^2(n)[hr_2(n)+hl_2(n)+ht_2(n)] \cr\cr
& & + 2f_2(n)[gt_2(n)gr_2(n)+gt_2(n)gl_2(n)+gr_2(n)gl_2(n)] \cr\cr
& & + f_2(n)[gr_2^2(n)+gl_2^2(n)+gt_2^2(n)] + 2gr_2(n)gl_2(n)gt_2(n) \cr\cr
& & + f_2(n)[gt_2(n)hr_2(n)+gl_2(n)ht_2(n)+gr_2(n)hl_2(n)] \cr\cr
& & + f_2(n)[ht_2(n)gr_2(n)+hl_2(n)gt_2(n)+hr_2(n)gl_2(n)] \ .
\eeqs
With the identity $gr_2(n)=gl_2(n)=gt_2(n)=g_2(n)$ and $hr_2(n)=hl_2(n)=ht_2(n)=h_2(n)$, Eq. (\ref{feq}) is verified.

\bigskip

\bigskip

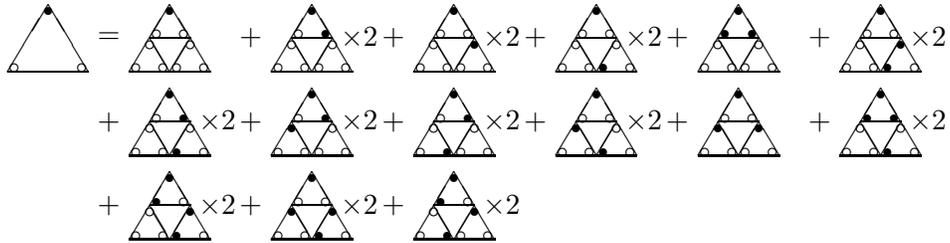
\begin{figure}[htbp]
\unitlength 0.9mm 
\begin{picture}(136,12)
\put(0,0){\line(1,0){12}}
\put(0,0){\line(3,5){6}}
\put(12,0){\line(-3,5){6}}
\multiput(1,0.5)(10,0){2}{\circle{1}}
\put(6,9){\circle*{1}}
\put(15,5){\makebox(0,0){$=$}}
\put(18,0){\line(1,0){12}}
\put(18,0){\line(3,5){6}}
\put(24,0){\line(3,5){3}}
\put(24,0){\line(-3,5){3}}
\put(21,5){\line(1,0){6}}
\put(30,0){\line(-3,5){6}}
\put(19,0.5){\circle{1}}
\put(21,4){\circle{1}}
\put(23,0.5){\circle{1}}
\put(22,5.5){\circle{1}}
\put(24,9){\circle*{1}}
\put(26,5.5){\circle{1}}
\put(25,0.5){\circle{1}}
\put(27,4){\circle{1}}
\put(29,0.5){\circle{1}}
\put(36,5){\makebox(0,0){$+$}}
\put(39,0){\line(1,0){12}}
\put(39,0){\line(3,5){6}}
\put(45,0){\line(3,5){3}}
\put(45,0){\line(-3,5){3}}
\put(42,5){\line(1,0){6}}
\put(51,0){\line(-3,5){6}}
\put(40,0.5){\circle{1}}
\put(42,4){\circle{1}}
\put(44,0.5){\circle{1}}
\put(43,5.5){\circle{1}}
\put(45,9){\circle*{1}}
\put(47,5.5){\circle*{1}}
\put(46,0.5){\circle{1}}
\put(48,4){\circle{1}}
\put(50,0.5){\circle{1}}
\put(52,5){\makebox(0,0){$\times 2$}}
\put(57,5){\makebox(0,0){$+$}}
\put(60,0){\line(1,0){12}}
\put(60,0){\line(3,5){6}}
\put(66,0){\line(3,5){3}}
\put(66,0){\line(-3,5){3}}
\put(63,5){\line(1,0){6}}
\put(72,0){\line(-3,5){6}}
\put(61,0.5){\circle{1}}
\put(63,4){\circle{1}}
\put(65,0.5){\circle{1}}
\put(64,5.5){\circle{1}}
\put(66,9){\circle*{1}}
\put(68,5.5){\circle{1}}
\put(67,0.5){\circle{1}}
\put(69,4){\circle*{1}}
\put(71,0.5){\circle{1}}
\put(73,5){\makebox(0,0){$\times 2$}}
\put(78,5){\makebox(0,0){$+$}}
\put(81,0){\line(1,0){12}}
\put(81,0){\line(3,5){6}}
\put(87,0){\line(3,5){3}}
\put(87,0){\line(-3,5){3}}
\put(84,5){\line(1,0){6}}
\put(93,0){\line(-3,5){6}}
\put(82,0.5){\circle{1}}
\put(84,4){\circle{1}}
\put(86,0.5){\circle{1}}
\put(85,5.5){\circle{1}}
\put(87,9){\circle*{1}}
\put(89,5.5){\circle{1}}
\put(88,0.5){\circle*{1}}
\put(90,4){\circle{1}}
\put(92,0.5){\circle{1}}
\put(94,5){\makebox(0,0){$\times 2$}}
\put(99,5){\makebox(0,0){$+$}}
\put(102,0){\line(1,0){12}}
\put(102,0){\line(3,5){6}}
\put(108,0){\line(3,5){3}}
\put(108,0){\line(-3,5){3}}
\put(105,5){\line(1,0){6}}
\put(114,0){\line(-3,5){6}}
\put(103,0.5){\circle{1}}
\put(105,4){\circle{1}}
\put(107,0.5){\circle{1}}
\put(106,5.5){\circle*{1}}
\put(108,9){\circle*{1}}
\put(110,5.5){\circle*{1}}
\put(109,0.5){\circle{1}}
\put(111,4){\circle{1}}
\put(113,0.5){\circle{1}}
\put(120,5){\makebox(0,0){$+$}}
\put(123,0){\line(1,0){12}}
\put(123,0){\line(3,5){6}}
\put(129,0){\line(3,5){3}}
\put(129,0){\line(-3,5){3}}
\put(126,5){\line(1,0){6}}
\put(135,0){\line(-3,5){6}}
\put(124,0.5){\circle{1}}
\put(126,4){\circle{1}}
\put(128,0.5){\circle{1}}
\put(127,5.5){\circle{1}}
\put(129,9){\circle*{1}}
\put(131,5.5){\circle{1}}
\put(130,0.5){\circle*{1}}
\put(132,4){\circle*{1}}
\put(134,0.5){\circle{1}}
\put(136,5){\makebox(0,0){$\times 2$}}
\end{picture}

\begin{picture}(136,12)
\put(15,5){\makebox(0,0){$+$}}
\put(18,0){\line(1,0){12}}
\put(18,0){\line(3,5){6}}
\put(24,0){\line(3,5){3}}
\put(24,0){\line(-3,5){3}}
\put(21,5){\line(1,0){6}}
\put(30,0){\line(-3,5){6}}
\put(19,0.5){\circle{1}}
\put(21,4){\circle{1}}
\put(23,0.5){\circle{1}}
\put(22,5.5){\circle{1}}
\put(24,9){\circle*{1}}
\put(26,5.5){\circle*{1}}
\put(25,0.5){\circle*{1}}
\put(27,4){\circle{1}}
\put(29,0.5){\circle{1}}
\put(31,5){\makebox(0,0){$\times 2$}}
\put(36,5){\makebox(0,0){$+$}}
\put(39,0){\line(1,0){12}}
\put(39,0){\line(3,5){6}}
\put(45,0){\line(3,5){3}}
\put(45,0){\line(-3,5){3}}
\put(42,5){\line(1,0){6}}
\put(51,0){\line(-3,5){6}}
\put(40,0.5){\circle{1}}
\put(42,4){\circle*{1}}
\put(44,0.5){\circle{1}}
\put(43,5.5){\circle{1}}
\put(45,9){\circle*{1}}
\put(47,5.5){\circle*{1}}
\put(46,0.5){\circle{1}}
\put(48,4){\circle{1}}
\put(50,0.5){\circle{1}}
\put(52,5){\makebox(0,0){$\times 2$}}
\put(57,5){\makebox(0,0){$+$}}
\put(60,0){\line(1,0){12}}
\put(60,0){\line(3,5){6}}
\put(66,0){\line(3,5){3}}
\put(66,0){\line(-3,5){3}}
\put(63,5){\line(1,0){6}}
\put(72,0){\line(-3,5){6}}
\put(61,0.5){\circle{1}}
\put(63,4){\circle{1}}
\put(65,0.5){\circle*{1}}
\put(64,5.5){\circle{1}}
\put(66,9){\circle*{1}}
\put(68,5.5){\circle*{1}}
\put(67,0.5){\circle{1}}
\put(69,4){\circle{1}}
\put(71,0.5){\circle{1}}
\put(73,5){\makebox(0,0){$\times 2$}}
\put(78,5){\makebox(0,0){$+$}}
\put(81,0){\line(1,0){12}}
\put(81,0){\line(3,5){6}}
\put(87,0){\line(3,5){3}}
\put(87,0){\line(-3,5){3}}
\put(84,5){\line(1,0){6}}
\put(93,0){\line(-3,5){6}}
\put(82,0.5){\circle{1}}
\put(84,4){\circle*{1}}
\put(86,0.5){\circle{1}}
\put(85,5.5){\circle{1}}
\put(87,9){\circle*{1}}
\put(89,5.5){\circle{1}}
\put(88,0.5){\circle*{1}}
\put(90,4){\circle{1}}
\put(92,0.5){\circle{1}}
\put(94,5){\makebox(0,0){$\times 2$}}
\put(99,5){\makebox(0,0){$+$}}
\put(102,0){\line(1,0){12}}
\put(102,0){\line(3,5){6}}
\put(108,0){\line(3,5){3}}
\put(108,0){\line(-3,5){3}}
\put(105,5){\line(1,0){6}}
\put(114,0){\line(-3,5){6}}
\put(103,0.5){\circle{1}}
\put(105,4){\circle*{1}}
\put(107,0.5){\circle{1}}
\put(106,5.5){\circle{1}}
\put(108,9){\circle*{1}}
\put(110,5.5){\circle{1}}
\put(109,0.5){\circle{1}}
\put(111,4){\circle*{1}}
\put(113,0.5){\circle{1}}
\put(120,5){\makebox(0,0){$+$}}
\put(123,0){\line(1,0){12}}
\put(123,0){\line(3,5){6}}
\put(129,0){\line(3,5){3}}
\put(129,0){\line(-3,5){3}}
\put(126,5){\line(1,0){6}}
\put(135,0){\line(-3,5){6}}
\put(124,0.5){\circle{1}}
\put(126,4){\circle{1}}
\put(128,0.5){\circle{1}}
\put(127,5.5){\circle*{1}}
\put(129,9){\circle*{1}}
\put(131,5.5){\circle*{1}}
\put(130,0.5){\circle*{1}}
\put(132,4){\circle{1}}
\put(134,0.5){\circle{1}}
\put(136,5){\makebox(0,0){$\times 2$}}
\end{picture}

\begin{picture}(136,12)
\put(15,5){\makebox(0,0){$+$}}
\put(18,0){\line(1,0){12}}
\put(18,0){\line(3,5){6}}
\put(24,0){\line(3,5){3}}
\put(24,0){\line(-3,5){3}}
\put(21,5){\line(1,0){6}}
\put(30,0){\line(-3,5){6}}
\put(19,0.5){\circle{1}}
\put(21,4){\circle{1}}
\put(23,0.5){\circle{1}}
\put(22,5.5){\circle*{1}}
\put(24,9){\circle*{1}}
\put(26,5.5){\circle{1}}
\put(25,0.5){\circle*{1}}
\put(27,4){\circle*{1}}
\put(29,0.5){\circle{1}}
\put(31,5){\makebox(0,0){$\times 2$}}
\put(36,5){\makebox(0,0){$+$}}
\put(39,0){\line(1,0){12}}
\put(39,0){\line(3,5){6}}
\put(45,0){\line(3,5){3}}
\put(45,0){\line(-3,5){3}}
\put(42,5){\line(1,0){6}}
\put(51,0){\line(-3,5){6}}
\put(40,0.5){\circle{1}}
\put(42,4){\circle*{1}}
\put(44,0.5){\circle{1}}
\put(43,5.5){\circle{1}}
\put(45,9){\circle*{1}}
\put(47,5.5){\circle{1}}
\put(46,0.5){\circle*{1}}
\put(48,4){\circle*{1}}
\put(50,0.5){\circle{1}}
\put(52,5){\makebox(0,0){$\times 2$}}
\put(57,5){\makebox(0,0){$+$}}
\put(60,0){\line(1,0){12}}
\put(60,0){\line(3,5){6}}
\put(66,0){\line(3,5){3}}
\put(66,0){\line(-3,5){3}}
\put(63,5){\line(1,0){6}}
\put(72,0){\line(-3,5){6}}
\put(61,0.5){\circle{1}}
\put(63,4){\circle{1}}
\put(65,0.5){\circle*{1}}
\put(64,5.5){\circle*{1}}
\put(66,9){\circle*{1}}
\put(68,5.5){\circle{1}}
\put(67,0.5){\circle{1}}
\put(69,4){\circle*{1}}
\put(71,0.5){\circle{1}}
\put(73,5){\makebox(0,0){$\times 2$}}
\end{picture}

\caption{\footnotesize{Illustration for the expression of $gt_2(n+1)$. The multiplication of two on the right-hand-side corresponds to the reflection symmetry with respect to the central vertical axis.}} 
\label{gfig}
\end{figure}

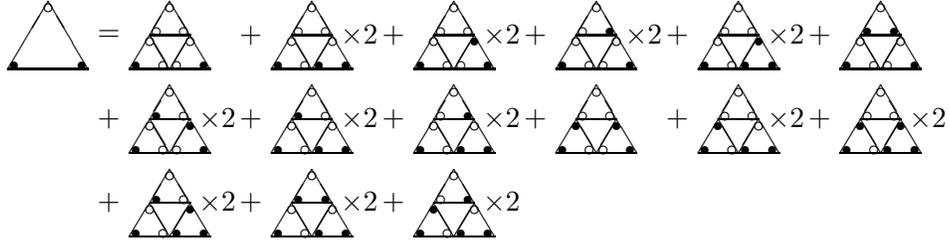
\begin{figure}[htbp]
\unitlength 0.9mm 
\begin{picture}(136,12)
\put(0,0){\line(1,0){12}}
\put(0,0){\line(3,5){6}}
\put(12,0){\line(-3,5){6}}
\multiput(1,0.5)(10,0){2}{\circle*{1}}
\put(6,9){\circle{1}}
\put(15,5){\makebox(0,0){$=$}}
\put(18,0){\line(1,0){12}}
\put(18,0){\line(3,5){6}}
\put(24,0){\line(3,5){3}}
\put(24,0){\line(-3,5){3}}
\put(21,5){\line(1,0){6}}
\put(30,0){\line(-3,5){6}}
\put(19,0.5){\circle*{1}}
\put(21,4){\circle{1}}
\put(23,0.5){\circle{1}}
\put(22,5.5){\circle{1}}
\put(24,9){\circle{1}}
\put(26,5.5){\circle{1}}
\put(25,0.5){\circle{1}}
\put(27,4){\circle{1}}
\put(29,0.5){\circle*{1}}
\put(36,5){\makebox(0,0){$+$}}
\put(39,0){\line(1,0){12}}
\put(39,0){\line(3,5){6}}
\put(45,0){\line(3,5){3}}
\put(45,0){\line(-3,5){3}}
\put(42,5){\line(1,0){6}}
\put(51,0){\line(-3,5){6}}
\put(40,0.5){\circle*{1}}
\put(42,4){\circle{1}}
\put(44,0.5){\circle{1}}
\put(43,5.5){\circle{1}}
\put(45,9){\circle{1}}
\put(47,5.5){\circle{1}}
\put(46,0.5){\circle*{1}}
\put(48,4){\circle{1}}
\put(50,0.5){\circle*{1}}
\put(52,5){\makebox(0,0){$\times 2$}}
\put(57,5){\makebox(0,0){$+$}}
\put(60,0){\line(1,0){12}}
\put(60,0){\line(3,5){6}}
\put(66,0){\line(3,5){3}}
\put(66,0){\line(-3,5){3}}
\put(63,5){\line(1,0){6}}
\put(72,0){\line(-3,5){6}}
\put(61,0.5){\circle*{1}}
\put(63,4){\circle{1}}
\put(65,0.5){\circle{1}}
\put(64,5.5){\circle{1}}
\put(66,9){\circle{1}}
\put(68,5.5){\circle{1}}
\put(67,0.5){\circle{1}}
\put(69,4){\circle*{1}}
\put(71,0.5){\circle*{1}}
\put(73,5){\makebox(0,0){$\times 2$}}
\put(78,5){\makebox(0,0){$+$}}
\put(81,0){\line(1,0){12}}
\put(81,0){\line(3,5){6}}
\put(87,0){\line(3,5){3}}
\put(87,0){\line(-3,5){3}}
\put(84,5){\line(1,0){6}}
\put(93,0){\line(-3,5){6}}
\put(82,0.5){\circle*{1}}
\put(84,4){\circle{1}}
\put(86,0.5){\circle{1}}
\put(85,5.5){\circle{1}}
\put(87,9){\circle{1}}
\put(89,5.5){\circle*{1}}
\put(88,0.5){\circle{1}}
\put(90,4){\circle{1}}
\put(92,0.5){\circle*{1}}
\put(94,5){\makebox(0,0){$\times 2$}}
\put(99,5){\makebox(0,0){$+$}}
\put(102,0){\line(1,0){12}}
\put(102,0){\line(3,5){6}}
\put(108,0){\line(3,5){3}}
\put(108,0){\line(-3,5){3}}
\put(105,5){\line(1,0){6}}
\put(114,0){\line(-3,5){6}}
\put(103,0.5){\circle*{1}}
\put(105,4){\circle{1}}
\put(107,0.5){\circle{1}}
\put(106,5.5){\circle{1}}
\put(108,9){\circle{1}}
\put(110,5.5){\circle{1}}
\put(109,0.5){\circle*{1}}
\put(111,4){\circle*{1}}
\put(113,0.5){\circle*{1}}
\put(115,5){\makebox(0,0){$\times 2$}}
\put(120,5){\makebox(0,0){$+$}}
\put(123,0){\line(1,0){12}}
\put(123,0){\line(3,5){6}}
\put(129,0){\line(3,5){3}}
\put(129,0){\line(-3,5){3}}
\put(126,5){\line(1,0){6}}
\put(135,0){\line(-3,5){6}}
\put(124,0.5){\circle*{1}}
\put(126,4){\circle{1}}
\put(128,0.5){\circle{1}}
\put(127,5.5){\circle*{1}}
\put(129,9){\circle{1}}
\put(131,5.5){\circle*{1}}
\put(130,0.5){\circle{1}}
\put(132,4){\circle{1}}
\put(134,0.5){\circle*{1}}
\end{picture}

\begin{picture}(136,12)
\put(15,5){\makebox(0,0){$+$}}
\put(18,0){\line(1,0){12}}
\put(18,0){\line(3,5){6}}
\put(24,0){\line(3,5){3}}
\put(24,0){\line(-3,5){3}}
\put(21,5){\line(1,0){6}}
\put(30,0){\line(-3,5){6}}
\put(19,0.5){\circle*{1}}
\put(21,4){\circle{1}}
\put(23,0.5){\circle{1}}
\put(22,5.5){\circle*{1}}
\put(24,9){\circle{1}}
\put(26,5.5){\circle{1}}
\put(25,0.5){\circle{1}}
\put(27,4){\circle*{1}}
\put(29,0.5){\circle*{1}}
\put(31,5){\makebox(0,0){$\times 2$}}
\put(36,5){\makebox(0,0){$+$}}
\put(39,0){\line(1,0){12}}
\put(39,0){\line(3,5){6}}
\put(45,0){\line(3,5){3}}
\put(45,0){\line(-3,5){3}}
\put(42,5){\line(1,0){6}}
\put(51,0){\line(-3,5){6}}
\put(40,0.5){\circle*{1}}
\put(42,4){\circle{1}}
\put(44,0.5){\circle{1}}
\put(43,5.5){\circle*{1}}
\put(45,9){\circle{1}}
\put(47,5.5){\circle{1}}
\put(46,0.5){\circle*{1}}
\put(48,4){\circle{1}}
\put(50,0.5){\circle*{1}}
\put(52,5){\makebox(0,0){$\times 2$}}
\put(57,5){\makebox(0,0){$+$}}
\put(60,0){\line(1,0){12}}
\put(60,0){\line(3,5){6}}
\put(66,0){\line(3,5){3}}
\put(66,0){\line(-3,5){3}}
\put(63,5){\line(1,0){6}}
\put(72,0){\line(-3,5){6}}
\put(61,0.5){\circle*{1}}
\put(63,4){\circle{1}}
\put(65,0.5){\circle{1}}
\put(64,5.5){\circle{1}}
\put(66,9){\circle{1}}
\put(68,5.5){\circle*{1}}
\put(67,0.5){\circle*{1}}
\put(69,4){\circle{1}}
\put(71,0.5){\circle*{1}}
\put(73,5){\makebox(0,0){$\times 2$}}
\put(78,5){\makebox(0,0){$+$}}
\put(81,0){\line(1,0){12}}
\put(81,0){\line(3,5){6}}
\put(87,0){\line(3,5){3}}
\put(87,0){\line(-3,5){3}}
\put(84,5){\line(1,0){6}}
\put(93,0){\line(-3,5){6}}
\put(82,0.5){\circle*{1}}
\put(84,4){\circle*{1}}
\put(86,0.5){\circle{1}}
\put(85,5.5){\circle{1}}
\put(87,9){\circle{1}}
\put(89,5.5){\circle{1}}
\put(88,0.5){\circle{1}}
\put(90,4){\circle*{1}}
\put(92,0.5){\circle*{1}}
\put(99,5){\makebox(0,0){$+$}}
\put(102,0){\line(1,0){12}}
\put(102,0){\line(3,5){6}}
\put(108,0){\line(3,5){3}}
\put(108,0){\line(-3,5){3}}
\put(105,5){\line(1,0){6}}
\put(114,0){\line(-3,5){6}}
\put(103,0.5){\circle*{1}}
\put(105,4){\circle*{1}}
\put(107,0.5){\circle{1}}
\put(106,5.5){\circle{1}}
\put(108,9){\circle{1}}
\put(110,5.5){\circle{1}}
\put(109,0.5){\circle*{1}}
\put(111,4){\circle{1}}
\put(113,0.5){\circle*{1}}
\put(115,5){\makebox(0,0){$\times 2$}}
\put(120,5){\makebox(0,0){$+$}}
\put(123,0){\line(1,0){12}}
\put(123,0){\line(3,5){6}}
\put(129,0){\line(3,5){3}}
\put(129,0){\line(-3,5){3}}
\put(126,5){\line(1,0){6}}
\put(135,0){\line(-3,5){6}}
\put(124,0.5){\circle*{1}}
\put(126,4){\circle*{1}}
\put(128,0.5){\circle{1}}
\put(127,5.5){\circle{1}}
\put(129,9){\circle{1}}
\put(131,5.5){\circle{1}}
\put(130,0.5){\circle*{1}}
\put(132,4){\circle*{1}}
\put(134,0.5){\circle*{1}}
\put(136,5){\makebox(0,0){$\times 2$}}
\end{picture}

\begin{picture}(136,12)
\put(15,5){\makebox(0,0){$+$}}
\put(18,0){\line(1,0){12}}
\put(18,0){\line(3,5){6}}
\put(24,0){\line(3,5){3}}
\put(24,0){\line(-3,5){3}}
\put(21,5){\line(1,0){6}}
\put(30,0){\line(-3,5){6}}
\put(19,0.5){\circle*{1}}
\put(21,4){\circle{1}}
\put(23,0.5){\circle{1}}
\put(22,5.5){\circle*{1}}
\put(24,9){\circle{1}}
\put(26,5.5){\circle{1}}
\put(25,0.5){\circle*{1}}
\put(27,4){\circle*{1}}
\put(29,0.5){\circle*{1}}
\put(31,5){\makebox(0,0){$\times 2$}}
\put(36,5){\makebox(0,0){$+$}}
\put(39,0){\line(1,0){12}}
\put(39,0){\line(3,5){6}}
\put(45,0){\line(3,5){3}}
\put(45,0){\line(-3,5){3}}
\put(42,5){\line(1,0){6}}
\put(51,0){\line(-3,5){6}}
\put(40,0.5){\circle*{1}}
\put(42,4){\circle{1}}
\put(44,0.5){\circle{1}}
\put(43,5.5){\circle*{1}}
\put(45,9){\circle{1}}
\put(47,5.5){\circle*{1}}
\put(46,0.5){\circle*{1}}
\put(48,4){\circle{1}}
\put(50,0.5){\circle*{1}}
\put(52,5){\makebox(0,0){$\times 2$}}
\put(57,5){\makebox(0,0){$+$}}
\put(60,0){\line(1,0){12}}
\put(60,0){\line(3,5){6}}
\put(66,0){\line(3,5){3}}
\put(66,0){\line(-3,5){3}}
\put(63,5){\line(1,0){6}}
\put(72,0){\line(-3,5){6}}
\put(61,0.5){\circle*{1}}
\put(63,4){\circle*{1}}
\put(65,0.5){\circle{1}}
\put(64,5.5){\circle{1}}
\put(66,9){\circle{1}}
\put(68,5.5){\circle*{1}}
\put(67,0.5){\circle*{1}}
\put(69,4){\circle{1}}
\put(71,0.5){\circle*{1}}
\put(73,5){\makebox(0,0){$\times 2$}}
\end{picture}

\caption{\footnotesize{Illustration for the expression of $ht_2(n+1)$. The multiplication of two on the right-hand-side corresponds to the reflection symmetry with respect to the central vertical axis.}} 
\label{hfig}
\end{figure}

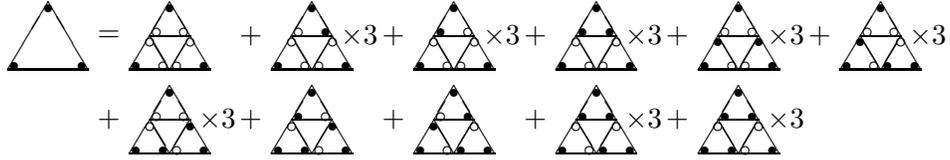
\begin{figure}[htbp]
\unitlength 0.9mm 
\begin{picture}(136,12)
\put(0,0){\line(1,0){12}}
\put(0,0){\line(3,5){6}}
\put(12,0){\line(-3,5){6}}
\multiput(1,0.5)(10,0){2}{\circle*{1}}
\put(6,9){\circle*{1}}
\put(15,5){\makebox(0,0){$=$}}
\put(18,0){\line(1,0){12}}
\put(18,0){\line(3,5){6}}
\put(24,0){\line(3,5){3}}
\put(24,0){\line(-3,5){3}}
\put(21,5){\line(1,0){6}}
\put(30,0){\line(-3,5){6}}
\put(19,0.5){\circle*{1}}
\put(21,4){\circle{1}}
\put(23,0.5){\circle{1}}
\put(22,5.5){\circle{1}}
\put(24,9){\circle*{1}}
\put(26,5.5){\circle{1}}
\put(25,0.5){\circle{1}}
\put(27,4){\circle{1}}
\put(29,0.5){\circle*{1}}
\put(36,5){\makebox(0,0){$+$}}
\put(39,0){\line(1,0){12}}
\put(39,0){\line(3,5){6}}
\put(45,0){\line(3,5){3}}
\put(45,0){\line(-3,5){3}}
\put(42,5){\line(1,0){6}}
\put(51,0){\line(-3,5){6}}
\put(40,0.5){\circle*{1}}
\put(42,4){\circle{1}}
\put(44,0.5){\circle{1}}
\put(43,5.5){\circle{1}}
\put(45,9){\circle*{1}}
\put(47,5.5){\circle*{1}}
\put(46,0.5){\circle{1}}
\put(48,4){\circle{1}}
\put(50,0.5){\circle*{1}}
\put(52,5){\makebox(0,0){$\times 3$}}
\put(57,5){\makebox(0,0){$+$}}
\put(60,0){\line(1,0){12}}
\put(60,0){\line(3,5){6}}
\put(66,0){\line(3,5){3}}
\put(66,0){\line(-3,5){3}}
\put(63,5){\line(1,0){6}}
\put(72,0){\line(-3,5){6}}
\put(61,0.5){\circle*{1}}
\put(63,4){\circle{1}}
\put(65,0.5){\circle{1}}
\put(64,5.5){\circle*{1}}
\put(66,9){\circle*{1}}
\put(68,5.5){\circle{1}}
\put(67,0.5){\circle{1}}
\put(69,4){\circle{1}}
\put(71,0.5){\circle*{1}}
\put(73,5){\makebox(0,0){$\times 3$}}
\put(78,5){\makebox(0,0){$+$}}
\put(81,0){\line(1,0){12}}
\put(81,0){\line(3,5){6}}
\put(87,0){\line(3,5){3}}
\put(87,0){\line(-3,5){3}}
\put(84,5){\line(1,0){6}}
\put(93,0){\line(-3,5){6}}
\put(82,0.5){\circle*{1}}
\put(84,4){\circle{1}}
\put(86,0.5){\circle{1}}
\put(85,5.5){\circle*{1}}
\put(87,9){\circle*{1}}
\put(89,5.5){\circle*{1}}
\put(88,0.5){\circle{1}}
\put(90,4){\circle{1}}
\put(92,0.5){\circle*{1}}
\put(94,5){\makebox(0,0){$\times 3$}}
\put(99,5){\makebox(0,0){$+$}}
\put(102,0){\line(1,0){12}}
\put(102,0){\line(3,5){6}}
\put(108,0){\line(3,5){3}}
\put(108,0){\line(-3,5){3}}
\put(105,5){\line(1,0){6}}
\put(114,0){\line(-3,5){6}}
\put(103,0.5){\circle*{1}}
\put(105,4){\circle*{1}}
\put(107,0.5){\circle{1}}
\put(106,5.5){\circle{1}}
\put(108,9){\circle*{1}}
\put(110,5.5){\circle{1}}
\put(109,0.5){\circle{1}}
\put(111,4){\circle*{1}}
\put(113,0.5){\circle*{1}}
\put(115,5){\makebox(0,0){$\times 3$}}
\put(120,5){\makebox(0,0){$+$}}
\put(123,0){\line(1,0){12}}
\put(123,0){\line(3,5){6}}
\put(129,0){\line(3,5){3}}
\put(129,0){\line(-3,5){3}}
\put(126,5){\line(1,0){6}}
\put(135,0){\line(-3,5){6}}
\put(124,0.5){\circle*{1}}
\put(126,4){\circle*{1}}
\put(128,0.5){\circle{1}}
\put(127,5.5){\circle{1}}
\put(129,9){\circle*{1}}
\put(131,5.5){\circle{1}}
\put(130,0.5){\circle*{1}}
\put(132,4){\circle{1}}
\put(134,0.5){\circle*{1}}
\put(136,5){\makebox(0,0){$\times 3$}}
\end{picture}

\begin{picture}(136,12)
\put(15,5){\makebox(0,0){$+$}}
\put(18,0){\line(1,0){12}}
\put(18,0){\line(3,5){6}}
\put(24,0){\line(3,5){3}}
\put(24,0){\line(-3,5){3}}
\put(21,5){\line(1,0){6}}
\put(30,0){\line(-3,5){6}}
\put(19,0.5){\circle*{1}}
\put(21,4){\circle{1}}
\put(23,0.5){\circle*{1}}
\put(22,5.5){\circle{1}}
\put(24,9){\circle*{1}}
\put(26,5.5){\circle{1}}
\put(25,0.5){\circle{1}}
\put(27,4){\circle*{1}}
\put(29,0.5){\circle*{1}}
\put(31,5){\makebox(0,0){$\times 3$}}
\put(36,5){\makebox(0,0){$+$}}
\put(39,0){\line(1,0){12}}
\put(39,0){\line(3,5){6}}
\put(45,0){\line(3,5){3}}
\put(45,0){\line(-3,5){3}}
\put(42,5){\line(1,0){6}}
\put(51,0){\line(-3,5){6}}
\put(40,0.5){\circle*{1}}
\put(42,4){\circle{1}}
\put(44,0.5){\circle*{1}}
\put(43,5.5){\circle*{1}}
\put(45,9){\circle*{1}}
\put(47,5.5){\circle{1}}
\put(46,0.5){\circle{1}}
\put(48,4){\circle*{1}}
\put(50,0.5){\circle*{1}}
\put(57,5){\makebox(0,0){$+$}}
\put(60,0){\line(1,0){12}}
\put(60,0){\line(3,5){6}}
\put(66,0){\line(3,5){3}}
\put(66,0){\line(-3,5){3}}
\put(63,5){\line(1,0){6}}
\put(72,0){\line(-3,5){6}}
\put(61,0.5){\circle*{1}}
\put(63,4){\circle*{1}}
\put(65,0.5){\circle{1}}
\put(64,5.5){\circle{1}}
\put(66,9){\circle*{1}}
\put(68,5.5){\circle*{1}}
\put(67,0.5){\circle*{1}}
\put(69,4){\circle{1}}
\put(71,0.5){\circle*{1}}
\put(78,5){\makebox(0,0){$+$}}
\put(81,0){\line(1,0){12}}
\put(81,0){\line(3,5){6}}
\put(87,0){\line(3,5){3}}
\put(87,0){\line(-3,5){3}}
\put(84,5){\line(1,0){6}}
\put(93,0){\line(-3,5){6}}
\put(82,0.5){\circle*{1}}
\put(84,4){\circle{1}}
\put(86,0.5){\circle{1}}
\put(85,5.5){\circle*{1}}
\put(87,9){\circle*{1}}
\put(89,5.5){\circle*{1}}
\put(88,0.5){\circle*{1}}
\put(90,4){\circle{1}}
\put(92,0.5){\circle*{1}}
\put(94,5){\makebox(0,0){$\times 3$}}
\put(99,5){\makebox(0,0){$+$}}
\put(102,0){\line(1,0){12}}
\put(102,0){\line(3,5){6}}
\put(108,0){\line(3,5){3}}
\put(108,0){\line(-3,5){3}}
\put(105,5){\line(1,0){6}}
\put(114,0){\line(-3,5){6}}
\put(103,0.5){\circle*{1}}
\put(105,4){\circle{1}}
\put(107,0.5){\circle*{1}}
\put(106,5.5){\circle*{1}}
\put(108,9){\circle*{1}}
\put(110,5.5){\circle*{1}}
\put(109,0.5){\circle{1}}
\put(111,4){\circle{1}}
\put(113,0.5){\circle*{1}}
\put(115,5){\makebox(0,0){$\times 3$}}
\end{picture}

\caption{\footnotesize{Illustration for the expression of $t_2(n+1)$. The multiplication of three on the right-hand-side corresponds to the three possible orientations of $SG_2(n+1)$.}} 
\label{tfig}
\end{figure}

Similarly, $gt_2(n+1)=g_2(n+1)$, $ht_2(n+1)=h_2(n+1)$ and $t_2(n+1)$ for $SG_2(n+1)$ can be obtained with appropriate configurations of its three constituting $SG_2(n)$ as illustrated in Figs. \ref{gfig}, \ref{hfig} and \ref{tfig} to verify Eqs. (\ref{geq}), (\ref{heq}) and (\ref{teq}), respectively. 

\bigskip

\bigskip

\bigskip

\bigskip

Eq. (\ref{Meq}) can also be obtained by substituting Eqs. (\ref{feq})-(\ref{teq}) into Eq. (\ref{Msg2}). \ $\Box$

\bigskip

There are always $27=3^3$ terms in Eqs. (\ref{feq}), (\ref{geq}), (\ref{heq}), (\ref{teq}) because there are three possible choices for each of the three pairs of identified vertices: both of them are originally occupied by monomers, or either one of them is originally occupied by a monomer while the other one by a dimer.
The values of $M_2(n)$, $f_2(n)$, $g_2(n)$, $h_2(n)$, $t_2(n)$ for small $n$ can be evaluated recursively by Eqs. (\ref{Meq})-(\ref{teq}) as listed in Table \ref{tablesg2}. These numbers grow exponentially, and do not have simple integer factorizations. To estimate the value of the asymptotic growth constant defined in Eq. (\ref{zdef}), we need the following lemmas. For the generalized two-dimensional Sierpinski gasket $SG_{2,b}(n)$, define the ratios
\beq
\alpha_{2,b}(n) = \frac{g_{2,b}(n)}{f_{2,b}(n)} \ , \qquad \beta_{2,b}(n) = \frac{h_{2,b}(n)}{g_{2,b}(n)} \ , \qquad \gamma_{2,b}(n) = \frac{t_{2,b}(n)}{h_{2,b}(n)} \ ,
\label{ratiodef}
\eeq
For the ordinary Sierpinski gasket in this section ,they are simplified to be $\alpha_2(n)$, $\beta_2(n)$ and $\gamma_2(n)$.

\bigskip

\begin{table}[htbp]
\caption{\label{tablesg2} The first few values of $M_2(n)$, $f_2(n)$, $g_2(n)$, $h_2(n)$, $t_2(n)$.}
\begin{center}
\begin{tabular}{|c||r|r|r|r|r|}
\hline\hline 
$n$      & 0 &  1 &      2 &               3 & 4 \\ \hline\hline 
$M_2(n)$ & 4 & 27 & 10,054 & 499,058,851,840 & 60,978,122,299,433,248,924,629,725,740,007,424 \\ \hline 
$f_2(n)$ & 1 &  4 &  1,584 &  78,721,368,064 & 9,618,673,427,679,675,357,952,788,786,053,120 \\ \hline 
$g_2(n)$ & 0 &  4 &  1,352 &  66,974,056,448 & 8,183,299,472,241,085,511,976,093,040,508,928 \\ \hline 
$h_2(n)$ & 1 &  3 &  1,148 &  56,979,607,552 & 6,962,123,286,110,084,944,276,569,997,705,216 \\ \hline 
$t_2(n)$ & 0 &  2 &    970 &  48,476,491,776 & 5,923,180,596,700,062,197,918,947,839,311,872 \\  \hline\hline 
\end{tabular}
\end{center}
\end{table}

\bigskip

\begin{lemma} \label{lemmasg2c} For any positive integer $n$, the magnitudes of $f_2(n)$, $g_2(n)$, $h_2(n)$, $t_2(n)$ are ordered as
\beq
t_2(n)\leq h_2(n)\leq g_2(n)\leq f_2(n) \ ,
\label{ordersg2}
\eeq
then
\beq
0 \le \gamma_2(n) \le \beta_2(n) \le \alpha_2(n) \le 1 \ .
\label{ratiosg2}
\eeq
\end{lemma}

{\sl Proof} \quad 
Eq. (\ref{ordersg2}) is valid for the first few positive integer $n$ by the numbers given in Table \ref{tablesg2}. By Eqs. (\ref{feq})-(\ref{teq}), we have
\beqs
\frac{f_2(n+1)}{f_2^3(n)} - \frac{g_2(n+1)}{f_2^2(n)g_2(n)} & = & [\alpha_2(n)-\beta_2(n)] [2+5\alpha_2(n)+2\alpha_2(n)\beta_2(n)] \cr\cr
& & + [1+2\alpha_2(n)][\alpha_2^2(n)-\beta_2(n)\gamma_2(n)] \ ,
\label{fmg}
\eeqs
\beqs
\frac{g_2(n+1)}{f_2^2(n)g_2(n)} - \frac{h_2(n+1)}{f_2(n)g_2^2(n)} & = & [\alpha_2(n)-\beta_2(n)] [2+3\alpha_2(n)+3\beta_2(n)+2\alpha_2(n)\beta_2(n)] \cr\cr
& & + \beta_2(n)[\alpha_2(n)-\gamma_2(n)+2[\alpha_2^2(n)-\beta_2(n)\gamma_2(n)]] \ ,
\label{gmh}
\eeqs
\beqs
\frac{h_2(n+1)}{f_2(n)g_2^2(n)} - \frac{t_2(n+1)}{g_2^3(n)} & = & [\alpha_2(n)-\beta_2(n)] [2+6\beta_2(n)+2\beta_2^2(n)+2\beta_2(n)\gamma_2(n)] \cr\cr
& & + \beta_2(n)[\alpha_2(n)-\gamma_2(n)][1+2\beta_2(n)] \ .
\label{hmt}
\eeqs
Eq. (\ref{ordersg2}) is proved by mathematical induction if Eqs. (\ref{fmg})-(\ref{hmt}) are larger or equal to zero. Equivalently, we need $\gamma_2(n) \le \beta_2(n) \le \alpha_2(n)$, which can be proved also by induction since the following two expressions are larger or equal to zero for any positive integer $n$.
\beqs
\lefteqn{\left[ \frac{g_2(n+1)}{f_2^2(n)g_2(n)} \right]^2 - \frac{f_2(n+1)}{f_2^3(n)} \frac{h_2(n+1)}{f_2(n)g_2^2(n)}} \cr\cr & = &
[\alpha_2(n)-\beta_2(n)]^2 [1 + 3\alpha_2(n)+\gamma_2(n) + 4\alpha_2^2(n)[\beta_2(n)+\gamma_2(n)] + 4\alpha_2^2(n)\beta_2^2(n)]
\cr\cr & & + [\alpha_2^2(n)-\beta_2^2(n)] [[\alpha_2(n)-\gamma_2(n)][1+2\alpha_2(n)\beta_2(n)] +2\alpha_2^2(n)[\beta_2^2(n)-\gamma_2^2(n)]]
\cr\cr & & + [\alpha_2(n)-\beta_2(n)] [4\alpha_2(n)[\alpha_2^2(n)-\beta_2(n)\gamma_2(n)] + \alpha_2(n)\beta_2(n)[\alpha_2(n)-\gamma_2(n)] \cr\cr & & + \alpha_2(n)\beta_2(n)[\beta_2(n)-\gamma_2(n)]] + 4\alpha_2(n)[\alpha_2^3(n)-\beta_2^3(n)][\beta_2(n)-\gamma_2(n)]
\cr\cr & & + [\alpha_2^2(n)-\beta_2(n)\gamma_2(n)]^2 + 2\alpha_2(n)[\beta_2(n)-\gamma_2(n)]^2 [2\beta_2^2(n) + \alpha_2^3(n)+\alpha_2(n)\beta_2^2(n)]  \ , \cr & &
\eeqs
\beqs
\lefteqn{\left[ \frac{h_2(n+1)}{f_2(n)g_2^2(n)} \right]^2 - \frac{g_2(n+1)}{f_2^2(n)g_2(n)} \frac{t_2(n+1)}{g_2^3(n)}} \cr\cr & = &
[\alpha_2(n)-\beta_2(n)]^2 [1 + 4\beta_2(n) + 2\gamma_2(n) + 9\beta_2^2(n) + 6\beta_2^3(n)+6\beta_2^2(n)\gamma_2(n) + 2\beta_2^2(n)\gamma_2^2(n)]
\cr\cr & & + [\alpha_2(n)-\beta_2(n)] [2\alpha_2(n)[\beta_2(n)-\gamma_2(n)] + 4\beta_2^2(n)[\alpha_2(n)+\beta_2(n)-2\gamma_2(n)] \cr\cr & & + 2\alpha_2(n)\beta_2^2(n)[\beta_2^2(n)-\gamma_2^2(n)] + 2\beta_2^3(n)[\alpha_2(n)\beta_2(n)-\gamma_2^2(n)]]
\cr\cr & & + \beta_2(n)[\alpha_2^2(n)-\beta_2^2(n)][\beta_2(n)-\gamma_2(n)] [1 + 8\beta_2(n)]
\cr\cr & & + \beta_2^2(n)[\beta_2(n)-\gamma_2(n)]^2 [1+ 2\alpha_2(n)+2\beta_2(n) +4\alpha_2^2(n)]  \ .
\eeqs
It is clear that $0 \le \gamma_2(n)$ since all the quantities $f_2(n)$, $g_2(n)$, $h_2(n)$, $t_2(n)$ are positive. Finally, $\alpha_2(n) \le 1$ once Eq. (\ref{ordersg2}) is established.
\ $\Box$

\bigskip

The values of $\alpha_2(n)$, $\beta_2(n)$, $\gamma_2(n)$ for small $n$ are listed in Table \ref{tablesg2n}. 

\bigskip

\begin{table}[htbp]
\caption{\label{tablesg2n} The first few values of $\alpha_2(n)$, $\beta_2(n)$, $\gamma_2(n)$. The last digits given are rounded off.}
\begin{center}
\begin{tabular}{|c||r|r|r|r|r|}
\hline\hline 
$n$           &                 1 &                 2 &                 3 &                    4 \\ \hline\hline 
$\alpha_2(n)$ &                 1 & 0.853535353535354 & 0.850773533223540 & 0.850772150002722 \\ \hline 
$\beta_2(n)$  &              0.75 & 0.849112426035503 & 0.850771337051088 & 0.850772150002159 \\ \hline 
$\gamma_2(n)$ & 0.666666666666667 & 0.844947735191638 & 0.850769141078411 & 0.850772150001597 \\ \hline\hline 
\end{tabular}
\end{center}
\end{table}

\bigskip

\begin{lemma} \label{lemmasg2d} Sequence $\{ \alpha_2(n) \}_{n=1}^{\infty}$ decreases monotonically, while sequences $\{ \gamma_2(n) \}_{n=1}^{\infty}$ and $\{ t_2(n)/f_2(n) \}_{n=1}^{\infty}$ increase monotonically. The limits $\alpha_2 \equiv \lim_{n \to \infty} \alpha_2(n)$, $\beta_2 \equiv \lim_{n \to \infty} \beta_2(n)$, $\gamma_2 \equiv \lim_{n \to \infty} \gamma_2(n)$ exist.
\end{lemma}

{\sl Proof} \quad 
By Eqs. (\ref{feq}) and (\ref{geq}), we have
\beqs
\lefteqn{f_2(n+1)g_2(n) - g_2(n+1)f_2(n)} \cr\cr & = & f_2^2(n)g_2(n) \Big [ [2f_2(n)+5g_2(n)+2h_2(n)] [\alpha_2(n)-\beta_2(n)] \cr\cr
& & + [f_2(n)+2g_2(n)][\frac{g_2^2(n)}{f_2^2(n)}-\frac{t_2(n)}{g_2(n)}] \Big ] \ge 0 \ .
\eeqs
By induction, $\alpha_2(n)$ decreases as positive $n$ increases using the results of Lemma \ref{lemmasg2c}. By Eqs. (\ref{heq}) and (\ref{teq}), we have
\beqs
\lefteqn{t_2(n+1)h_2(n) - h_2(n+1)t_2(n)} \cr\cr & = & f_2(n)h_2(n)[g_2^2(n)+4g_2(n)h_2(n)+3h_2^2(n)] [\alpha_2(n)-\gamma_2(n)] \cr\cr
& & + 2g_2^2(n)h_2(n)[g_2(n)+2h_2(n)+t_2(n)] [\beta_2(n)-\gamma_2(n)]
\cr\cr & & + 2f_2(n)h_2^2(n)[g_2(n)+h_2(n)][\frac{h_2(n)}{f_2(n)}-\frac{t_2^2(n)}{h_2^2(n)}] \ge 0 \ ,
\eeqs
such that $\gamma_2(n)$ increases as positive $n$ increases. Finally, we have
\beqs
\lefteqn{t_2(n+1)f_2(n) - f_2(n+1)t_2(n)} \cr\cr & = & f_2^4(n)[\frac{g_2^3(n)}{f_2^3(n)} - \frac{t_2(n)}{f_2(n)}] + 6f_2^2(n)g_2(n)h_2(n)[\alpha_2(n)-\gamma_2(n)] \cr\cr
& & + 3f_2^2(n)g_2(n)t_2(n)[\alpha_2(n)-\beta_2(n)] + 9f_2(n)g_2^2(n)h_2(n)[\beta_2(n)-\gamma_2(n)] \cr\cr
& & - 2f_2(n)g_2^2(n)t_2(n)[\frac{g_2(n)}{f_2(n)}-\frac{h_2^3(n)}{g_2^2(n)t_2(n)}] \ge 0 \ ,
\eeqs
where the last inequality holds because of the combination of the third and the last terms:
\beq
2f_2(n)g_2(n)t_2(n) \left[ f_2(n)[\frac{g_2(n)}{f_2(n)}-\frac{h_2(n)}{g_2(n)}] - g_2(n)[\frac{g_2(n)}{f_2(n)}-\frac{h_2^3(n)}{g_2^2(n)t_2(n)}] \right] \ge 0 \ .
\eeq

Because the sequence $\alpha_2(n)$ decreases monotonically and bounded below, the limit $\alpha_2$ exists. Similarly, sequence $\gamma_2(n)$ and $t_2(n)/f_2(n)$ increases monotonically and bounded above so that the limits $\gamma_2$ and $\lim_{n \to \infty} t_2(n)/f_2(n)$ exist. It follows that the limit $\lim_{n \to \infty} h_2(n)/f_2(n)$ exists since
\beq
\lim_{n \to \infty} \frac{t_2(n)}{f_2(n)} = \lim_{n \to \infty} \frac{h_2(n)}{f_2(n)} \lim_{n \to \infty} \frac{t_2(n)}{h_2(n)} \ ,
\eeq
such that $\beta = \lim_{n \to \infty} h_2(n)/g_2(n)$ exists because
\beq
\lim_{n \to \infty} \frac{h_2(n)}{f_2(n)} = \lim_{n \to \infty} \frac{g_2(n)}{f_2(n)} \lim_{n \to \infty} \frac{h_2(n)}{g_2(n)} \ .
\eeq
\ $\Box$

\bigskip

With the existence of the limits $\alpha_2$, $\beta_2$, $\gamma_2$, and $\gamma_2 \le \beta_2 \le \alpha_2$, we have
\beqs
1 & = & \lim_{n \to \infty} \frac{f_2(n+1)}{f_2(n)} \frac{g_2(n)}{g_2(n+1)} \cr\cr 
& = & \frac {(1+3\alpha_2)^2+2\alpha_2^3+3\alpha_2\beta_2(1+2\alpha_2)} {1+2\beta_2+4\alpha_2+\beta_2\gamma_2+8\alpha_2\beta_2+3\alpha_2^2+2\alpha_2\beta_2\gamma_2+2\alpha_2\beta_2^2+4\alpha_2^2\beta_2}
\eeqs
by Eqs. (\ref{feq}) and (\ref{geq}), which leads to the following result.

\bigskip

\begin{cor} \label{corsg2} The three limits $\gamma_2$, $\beta_2$ and $\alpha_2$ are equal to each other.
\end{cor}

\bigskip

\begin{lemma} \label{lemmasg2b} The asymptotic growth constant for the number of dimer-monomers on $SG_2(n)$ is bounded:
\beq
\frac{2}{3^{m+1}} \ln f_2(m) + \frac{\ln[1+2\gamma_2(m)]}{3^m} \le z_{SG_2} \le \frac{2}{3^{m+1}} \ln f_2(m) + \frac{\ln[1+2\alpha_2(m)]}{3^m} \ ,
\label{zsg2}
\eeq
where $m$ is a positive integer.
\end{lemma}

{\sl Proof} \quad 
Let us define $\lambda_2(n) = f_2(n+1)/f_2^3(n)$. By Eq. (\ref{feq}), we have
\beq
\lambda_2(n) = [1+3\alpha_2(n)]^2 + 2\alpha_2^3(n) + 3\alpha_2(n)\beta_2(n)[1+2\alpha_2(n)] \ .
\eeq
It is clear that $1\leq \lambda_n\leq 27$, and
\beq
[1+2\gamma_2(m)]^3 \leq [1+2\gamma_2(n)]^3 \leq [1+2\beta_2(n)]^3 \leq \lambda_2(n) \leq [1+2\alpha_2(n)]^3 \leq [1+2\alpha_2(m)]^3
\label{lambdainequality}
\eeq
for $n \ge m$. By Eqs. (\ref{v}) and (\ref{Msg2}), we have
\beq
\frac {\ln M_2(n)}{v(SG_2(n))} = \frac {2\ln [1+3\alpha_2(n)+3\alpha_2(n)\beta_2(n)+\alpha_2(n)\beta_2(n)\gamma_2(n)]}{3(3^n+1)}+\frac {2\ln f_2(n)}{3(3^n+1)} \ ,
\eeq
where
\beqs
\ln f_2(n) & = & \ln \lambda_2(n-1)+3\ln f_2(n-1) \cr\cr
& = & \ln \lambda_2(n-1)+3\ln \lambda_2(n-2)+3^2\ln f_2(n-2) \cr\cr
& = & \cdots \cr\cr
& = & \sum_{j=m}^{n-1} 3^{n-1-j} \ln \lambda_2(j)+3^{n-m}\ln f_2(m)
\eeqs
for any $m < n$. By the definition of the asymptotic growth constant in Eq. (\ref{zdef}),
\beqs
z_{SG_2} & = & \lim_{n \to \infty} \frac{\ln M_2(n)}{v(SG_2(n))} \cr\cr
& = & \lim_{n\to\infty}
\frac{2\ln [1+3\alpha_2(n)+3\alpha_2(n)\beta_2(n)+\alpha_2(n)\beta_2(n)\gamma_2(n)]}{3(3^n+1)} \cr\cr
& & +\lim_{n\to\infty} \frac{ 2\sum_{j=m}^{n-1} 3^{n-1-j}\ln \lambda_2(j) + 2[3^{n-m}\ln f_2(m)]}{3(3^n+1)} \cr\cr
& = & \frac {2}{3^{2}}\sum_{j=m}^{\infty} \frac{\ln \lambda_2(j)}{3^j} + \frac{2}{3^{m+1}}\ln f_2(m) \ .
\eeqs
The proof is completed using the inequality (\ref{lambdainequality}).
\ $\Box$

\bigskip

The difference between the upper and lower bounds for $z_{SG_2}$ quickly converges to zero as $m$ increases, and we have the following proposition.

\bigskip

\begin{propo} \label{proposg2} The asymptotic growth constant for the number of dimer-monomers on the two-dimensional Sierpinski gasket $SG_2(n)$ in the large $n$ limit is $z_{SG_2}=0.656294236916...$. 

\end{propo}

\bigskip

The numerical value of $z_{SG_2}$ can be calculated with more than a hundred significant figures accurate when $m$ in Eq. (\ref{zsg2}) is equal to seven.

\section{The number of dimer-monomers on $SG_{2,b}(n)$ with $b=3,4$} 
\label{sectionIV}

The method given in the previous section can be applied to the number of dimer-monomers on $SG_{d,b}(n)$ with larger values of $d$ and $b$. The number of configurations to be considered increases as $d$ and $b$ increase, and the recursion relations must be derived individually for each $d$ and $b$. 
In this section, we consider the generalized two-dimensional Sierpinski gasket $SG_{2,b}(n)$ with the number of layers $b$ equal to three and four. 
For $SG_{2,3}(n)$, the numbers of edges and vertices are given by 
\beq
e(SG_{2,3}(n)) = 3 \times 6^n \ ,
\label{esg23}
\eeq
\beq
v(SG_{2,3}(n)) = \frac{7 \times 6^n + 8}{5} \ ,
\label{vsg23}
\eeq
where the three outmost vertices have degree two. There are $(6^n-1)/5$ vertices of $SG_{2,3}(n)$ with degree six and $6(6^n-1)/5$ vertices with degree four. By Definition \ref{defisg2}, the number of dimer-monomers is $M_{2,3}(n) = f_{2,3}(n)+3g_{2,3}(n)+3h_{2,3}(n)+t_{2,3}(n)$. The initial values are the same as for $SG_2$: $f_{2,3}(0)=1$, $g_{2,3}(0)=0$, $h_{2,3}(0)=1$ and $t_{2,3}(0)=0$. 

The recursion relations are lengthy and given in the appendix.
Some values of $M_{2,3}(n)$, $f_{2,3}(n)$, $g_{2,3}(n)$, $h_{2,3}(n)$, $t_{2,3}(n)$ are listed in Table \ref{tablesg23}. These numbers grow exponentially, and do not have simple integer factorizations.

\bigskip

\begin{table}[htbp]
\caption{\label{tablesg23} The first few values of $M_{2,3}(n)$, $f_{2,3}(n)$, $g_{2,3}(n)$, $h_{2,3}(n)$, $t_{2,3}(n)$.}
\begin{center}
\begin{tabular}{|c||r|r|r|}
\hline\hline 
$n$          & 0 &   1 &                   2 \\ \hline\hline 
$M_{2,3}(n)$ & 4 & 425 & 755,290,432,490,932 \\  \hline 
$f_{2,3}(n)$ & 1 &  66 & 116,464,644,336,176 \\  \hline 
$g_{2,3}(n)$ & 0 &  56 & 100,722,462,529,064 \\ \hline 
$h_{2,3}(n)$ & 1 &  49 &  87,108,127,443,640 \\ \hline 
$t_{2,3}(n)$ & 0 &  44 &  75,334,018,236,644 \\
\hline\hline 
\end{tabular}
\end{center}
\end{table}

\bigskip

The sequence of the ratio defined in Eq. (\ref{ratiodef}) $\{\alpha_{2,3}(n)\}_{n=1}^\infty$ increases monotonically and $\{\gamma_{2,3}(n)\}_{n=1}^\infty$ decreases monotonically with $0 \le \alpha_{2,3}(n) \le \gamma_{2,3}(n) \le 1$, in contrast to the results for $SG_2(n)$. 
The values of $\alpha_{2,3}(n)$, $\beta_{2,3}(n)$, $\gamma_{2,3}(n)$ for small $n$ are listed in Table \ref{tablesg23n}.

\bigskip

\begin{table}[htbp]
\caption{\label{tablesg23n} The first few values of $\alpha_{2,3}(n)$, $\beta_{2,3}(n)$, $\gamma_{2,3}(n)$. The last digits given are rounded off.}
\begin{center}
\begin{tabular}{|c||r|r|r|}
\hline\hline 
$n$               &                 1 &                 2 &                               3 \\ \hline\hline 
$\alpha_{2,3}(n)$ & 0.848484848484848 & 0.864832955126948 & 0.864833096846111 \\ \hline 
$\beta_{2,3}(n)$  &             0.875 & 0.864833178780796 & 0.864833096846111 \\ \hline 
$\gamma_{2,3}(n)$ & 0.897959183673469 & 0.864833402432925 & 0.864833096846111 \\ \hline\hline 
\end{tabular}
\end{center}
\end{table}

\bigskip

By a similar argument as Lemma \ref{lemmasg2b}, the asymptotic growth constant for the number of dimer-monomers on $SG_{2,3}(n)$ is bounded:
\beq \textstyle
\frac{5\ln f_{2,3}(m)+6\ln [1+2\alpha_{2,3}(m)]+\ln [1+3\alpha_{2,3}(m)]}{7\times 6^m} \le z_{SG_{2,3}} \le \frac{5\ln f_{2,3}(m)+6\ln [1+2\gamma_{2,3}(m)]+\ln [1+3\gamma_{2,3}(m)]}{7\times 6^m} \ ,
\label{zsg23}
\eeq
with $m$ a positive integer. We have the following proposition.

\bigskip

\begin{propo} \label{proposg23} The asymptotic growth constant for the number of dimer-monomers on the generalized two-dimensional Sierpinski gasket $SG_{2,3}(n)$ in the large $n$ limit is $z_{SG_{2,3}}=0.671617161058...$.

\end{propo}

\bigskip

The convergence of the upper and lower bounds remains quick. More than a hundred significant figures for $z_{SG_{2,3}}$ can be obtained when $m$  in Eq. (\ref{zsg23}) is equal to five.

For $SG_{2,4}(n)$, the numbers of edges and vertices are given by 
\beq
e(SG_{2,4}(n)) = 3 \times 10^n \ ,
\label{esg24}
\eeq
\beq
v(SG_{2,4}(n)) = \frac{4 \times 10^n + 5}{3} \ ,
\label{vsg24}
\eeq
where again the three outmost vertices have degree two. There are $(10^n-1)/3$ vertices of $SG_{2,4}(n)$ with degree six, and $(10^n-1)$ vertices with degree four. By Definition \ref{defisg2}, the number of dimer-monomers is $M_{2,4}(n) = f_{2,4}(n)+3g_{2,4}(n)+3h_{2,4}(n)+t_{2,4}(n)$. The initial values are the same as for $SG_2$: $f_{2,4}(0)=1$, $g_{2,4}(0)=0$, $h_{2,4}(0)=1$ and $t_{2,4}(0)=0$.
We write a computer program to obtain the recursion relations for $SG_{2,4}(n)$. They are too lengthy to be included here and are available from the authors on request. Some values of $M_{2,4}(n)$, $f_{2,4}(n)$, $g_{2,4}(n)$, $h_{2,4}(n)$, $t_{2,4}(n)$ are listed in Table \ref{tablesg24}. These numbers grow exponentially, and do not have simple integer factorizations.

\bigskip

\begin{table}[htbp]
\caption{\label{tablesg24} The first few values of $M_{2,4}(n)$, $f_{2,4}(n)$, $g_{2,4}(n)$, $h_{2,4}(n)$, $t_{2,4}(n)$.}
\begin{center}
\begin{tabular}{|c||r|r|r|}
\hline\hline 
$n$          & 0 & 1      & 2 \\ \hline \hline
$M_{2,4}(n)$ & 4 & 14,278 & 7,033,761,314,434,948,243,456,944,474,554,222,281,728 \\ \hline 
$f_{2,4}(n)$ & 1 &  2,220 & 1,095,249,688,634,151,454,219,516,689,432,826,798,080 \\ \hline 
$g_{2,4}(n)$ & 0 &  1,914 &  940,563,707,718,765,231,855,988,194,853,818,067,968 \\ \hline 
$h_{2,4}(n)$ & 1 &  1,640 &  807,724,574,091,886,425,362,687,789,454,449,995,776 \\ \hline 
$t_{2,4}(n)$ & 0 &  1,396 &  693,646,780,368,841,817,581,399,832,196,591,292,416 \\ \hline\hline 
\end{tabular}
\end{center}
\end{table}

\bigskip

The sequence of the ratio defined in Eq. (\ref{ratiodef}) $\{\alpha_{2,4}(n)\}_{n=1}^\infty$ decreases monotonically and $\{\gamma_{2,4}(n)\}_{n=1}^\infty$ increases monotonically with $0 \le \gamma_{2,4}(n) \le \alpha_{2,4}(n) \le 1$, the same as the results for $SG_2(n)$.

The values of $\alpha_{2,4}(n)$, $\beta_{2,4}(n)$, $\gamma_{2,4}(n)$ for small $n$ are listed in Table \ref{tablesg24n}.

\bigskip

\begin{table}[htbp]
\caption{\label{tablesg24n} The first few values of $\alpha_{2,4}(n)$, $\beta_{2,4}(n)$, $\gamma_{2,4}(n)$. The last digits given are rounded off.}
\begin{center}
\begin{tabular}{|c||r|r|r|}
\hline\hline 
$n$               &                 1 &                 2 &                               3 \\ \hline\hline 
$\alpha_{2,4}(n)$ & 0.862162162162162 & 0.858766468942539 & 0.858766468941692 \\ \hline 
$\beta_{2,4}(n)$  & 0.856844305120167 & 0.858766468941199 & 0.858766468941692 \\ \hline 
$\gamma_{2,4}(n)$ & 0.851219512195122 & 0.858766468939860 & 0.858766468941692 \\ \hline\hline 
\end{tabular}
\end{center}
\end{table}

\bigskip

By a similar argument as Lemma \ref{lemmasg2b}, the asymptotic growth constant for the number of dimer-monomers on $SG_{2,4}(n)$ is bounded:
\beq \textstyle
\frac{3\ln f_{2,4}(m)+3\ln [1+2\gamma_{2,4}(m)]+\ln [1+3\gamma_{2,4}(m)]}{4\times 10^m} \le z_{SG_{2,4}} \le \frac{3\ln f_{2,4}(m)+3\ln [1+2\alpha_{2,4}(m)]+\ln [1+3\alpha_{2,4}(m)]}{4\times 10^m} \ ,
\label{zsg24}
\eeq
with $m$ a positive integer. We have the following proposition.

\bigskip

\begin{propo} \label{proposg24} The asymptotic growth constant for the number of dimer-monomers on the generalized two-dimensional Sierpinski gasket $SG_{2,4}(n)$ in the large $n$ limit is $z_{SG_{2,4}}=0.684872262332...$.

\end{propo}

\bigskip

The convergence of the upper and lower bounds is again quick. 
More than a hundred significant figures for $z_{SG_{2,4}}$ can be obtained when $m$ in Eq. (\ref{zsg24}) is equal to four.

\section{The number of dimer-monomers on $SG_d(n)$ with $d=3,4$} 
\label{sectionV}

In this section, we derive the asymptotic growth constants of dimer-monomers on $SG_d(n)$ with $d=3,4$.
For the three-dimensional Sierpinski gasket $SG_3(n)$, we use the following definitions.

\bigskip

\begin{defi} \label{defisg3} Consider the three-dimensional Sierpinski gasket $SG_3(n)$ at stage $n$. (a) Define $M_3(n) \equiv N_{DM}(SG_3(n))$ as the number of dimer-monomers. (b) Define $f_3(n)$ as the number of dimer-monomers such that the four outmost vertices are occupied by monomers. (c) Define $g_3(n)$ as the number of dimer-monomers such that one of the outmost vertices is occupied by a dimer and the other three outmost vertices are occupied by monomers. (d) Define $h_3(n)$ as the number of dimer-monomers such that two of the outmost vertices are occupied by monomers and the other two outmost vertices are occupied by dimers. (e) Define $r_3(n)$ as the number of dimer-monomers such that one of the outmost vertices is occupied by a monomer and the other three outmost vertices are occupied by dimers. (f) Define $s_3(n)$ as the number of dimer-monomers such that all four outmost vertices are occupied by dimers.
\end{defi}

\bigskip

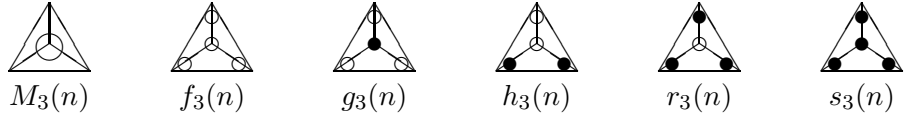
\begin{figure}[htbp]
\unitlength 1.8mm 
\begin{picture}(66,5)
\put(0,0){\line(1,0){6}}
\put(0,0){\line(3,5){3}}
\put(6,0){\line(-3,5){3}}
\put(0,0){\line(3,2){3}}
\put(6,0){\line(-3,2){3}}
\put(3,2){\line(0,1){3}}
\put(3,1.7){\circle{2}}
\put(3,-2){\makebox(0,0){$M_3(n)$}}
\put(12,0){\line(1,0){6}}
\put(12,0){\line(3,5){3}}
\put(18,0){\line(-3,5){3}}
\put(12,0){\line(3,2){3}}
\put(18,0){\line(-3,2){3}}
\put(15,2){\line(0,1){3}}
\put(13,0.5){\circle{1}}
\put(17,0.5){\circle{1}}
\put(15,2){\circle{1}}
\put(15,4){\circle{1}}
\put(15,-2){\makebox(0,0){$f_3(n)$}}
\put(24,0){\line(1,0){6}}
\put(24,0){\line(3,5){3}}
\put(30,0){\line(-3,5){3}}
\put(24,0){\line(3,2){3}}
\put(30,0){\line(-3,2){3}}
\put(27,2){\line(0,1){3}}
\put(25,0.5){\circle{1}}
\put(29,0.5){\circle{1}}
\put(27,2){\circle*{1}}
\put(27,4){\circle{1}}
\put(27,-2){\makebox(0,0){$g_3(n)$}}
\put(36,0){\line(1,0){6}}
\put(36,0){\line(3,5){3}}
\put(42,0){\line(-3,5){3}}
\put(36,0){\line(3,2){3}}
\put(42,0){\line(-3,2){3}}
\put(39,2){\line(0,1){3}}
\put(37,0.5){\circle*{1}}
\put(41,0.5){\circle*{1}}
\put(39,2){\circle{1}}
\put(39,4){\circle{1}}
\put(39,-2){\makebox(0,0){$h_3(n)$}}
\put(48,0){\line(1,0){6}}
\put(48,0){\line(3,5){3}}
\put(54,0){\line(-3,5){3}}
\put(48,0){\line(3,2){3}}
\put(54,0){\line(-3,2){3}}
\put(51,2){\line(0,1){3}}
\put(49,0.5){\circle*{1}}
\put(53,0.5){\circle*{1}}
\put(51,2){\circle{1}}
\put(51,4){\circle*{1}}
\put(51,-2){\makebox(0,0){$r_3(n)$}}
\put(60,0){\line(1,0){6}}
\put(60,0){\line(3,5){3}}
\put(66,0){\line(-3,5){3}}
\put(60,0){\line(3,2){3}}
\put(66,0){\line(-3,2){3}}
\put(63,2){\line(0,1){3}}
\put(61,0.5){\circle*{1}}
\put(65,0.5){\circle*{1}}
\put(63,2){\circle*{1}}
\put(63,4){\circle*{1}}
\put(63,-2){\makebox(0,0){$s_3(n)$}}
\end{picture}

\vspace*{5mm}
\caption{\footnotesize{Illustration for the spanning subgraphs $M_3(n)$, $f_3(n)$, $g_3(n)$, $h_3(n)$, $r_3(n)$ and $s_3(n)$. Only the four outmost vertices are shown explicitly for $f_3(n)$, $g_3(n)$, $h_3(n)$, $r_3(n)$ and $s_3(n)$, where each open circle is occupied by a monomer and each solid circle is occupied by a dimer.}} 
\label{fghrsfig}
\end{figure}

The quantities $M_3(n)$, $f_3(n)$, $g_3(n)$, $h_3(n)$, $r_3(n)$ and $s_3(n)$ are illustrated in Fig. \ref{fghrsfig}, where only the outmost vertices are shown. There are ${4 \choose 1}=4$ equivalent $g_3(n)$, ${4 \choose 2}=6$ equivalent $h_3(n)$, and ${4 \choose 1}=4$ equivalent $r_3(n)$. By definition,
\beq
M_3(n) = f_3(n)+4g_3(n)+6h_3(n)+4r_3(n)+s_3(n) \ .
\label{fsg3}
\eeq
The initial values at stage zero are $f_3(0)=1$, $g_3(0)=0$, $h_3(0)=1$, $r_3(0)=0$, $s_3(0)=3$ and $M_3(0)=10$. 

\bigskip

\bigskip

The recursion relations are lengthy and given in the appendix. Some values of $M_3(n)$, $f_3(n)$, $g_3(n)$, $h_3(n)$, $r_3(n)$, $s_3(n)$ are listed in Table \ref{tablesg3}. These numbers grow exponentially, and do not have simple integer factorizations.

\bigskip

\begin{table}[htbp]
\caption{\label{tablesg3} The first few values of $M_3(n)$, $f_3(n)$, $g_3(n)$, $h_3(n)$, $r_3(n)$, $s_3(n)$.}
\begin{center}
\begin{tabular}{|c||r|r|r|r|}
\hline\hline 
$n$      & 0  &   1 &               2 & 3 \\ \hline\hline 
$M_3(n)$ & 10 & 945 & 132,820,373,046 & 49,123,375,811,021,432,878,640,796,802,876,545,882,185,505 \\ \hline 
$f_3(n)$ &  1 &  51 &   7,365,569,811 & 2,724,928,560,954,289,860,903,291,271,266,882,549,492,483 \\ \hline 
$g_3(n)$ &  0 &  57 &   7,816,070,424 & 2,889,924,536,764,017,260,444,663,495,693,780,813,791,233 \\ \hline 
$h_3(n)$ &  1 &  62 &   8,289,450,499 & 3,064,910,998,294,837,201,844,707,724,238,032,710,560,958 \\ \hline
$r_3(n)$ &  0 &  60 &   8,786,476,992 & 3,250,492,861,272,219,038,243,497,885,127,347,116,333,900 \\ \hline 
$s_3(n)$ &  3 &  54 &   9,307,910,577 & 3,447,311,668,153,174,611,916,613,662,896,955,348,826,742 \\ \hline\hline 
\end{tabular}
\end{center}
\end{table}

\bigskip

Define $\alpha_3(n)=g_3(n)/f_3(n)$ and $\gamma_3(n)=s_3(n)/r_3(n)$ as in Eq. (\ref{ratiodef}). 
We find $\{\alpha_3(n)\}_{n=1}^\infty$ decreases monotonically and $\{\gamma_3(n)\}_{n=1}^\infty$ increases monotonically with $1 \le \gamma_3(n) \le \alpha_3(n)$ for $n \ge 2$.
The values of $\alpha_3(n)$, $\gamma_3(n)$ and other ratios for small $n$ are listed in Table \ref{tablesg3n}. 

\bigskip

\begin{table}[htbp]
\caption{\label{tablesg3n} The first few values of $\alpha_3(n)$, $\gamma_3(n)$ and other ratios. The last digits given are rounded off.}
\begin{center}
\begin{tabular}{|c||r|r|r|r|}
\hline\hline 
$n$              &                1 &                2 &                3 & 4 \\ \hline\hline 
$\alpha_3(n)$    & 1.11764705882353 & 1.06116303620220 & 1.06055056935215 & 1.06055052894365 \\ \hline 
$h_3(n)/g_3(n)$  & 1.08771929824561 & 1.06056497054408 & 1.06055052971271 & 1.06055052894365 \\ \hline 
$r_3(n)/h_3(n)$  & 0.96774193548387 & 1.05995891923837 & 1.06055049007316 & 1.06055052894365 \\ \hline 
$\gamma_3(n)$    &              0.9 & 1.05934501228135 & 1.06055045043351 & 1.06055052894365 \\ \hline\hline 
\end{tabular}
\end{center}
\end{table}

\bigskip

By a similar argument as Lemma \ref{lemmasg2b}, the asymptotic growth constant for the number of dimer-monomers on $SG_3(n)$ is bounded:
\beq
\frac{\ln f_3(m) +2\ln [1+2\gamma_3(m)]}{2\times 4^m}  \le z_{SG_3} \le \frac{\ln f_3(m) +2 \ln [1+2\alpha_3(m)]}{2\times 4^m}   \ ,
\label{zsg3}
\eeq
with $m$ a positive integer. We have the following proposition.

\bigskip

\begin{propo} \label{proposg3} The asymptotic growth constant for the number of dimer-monomers on the three-dimensional Sierpinski gasket $SG_3(n)$ in the large $n$ limit is $z_{SG_3}=0.781151467411...$.

\end{propo}

\bigskip

The convergence of the upper and lower bounds is as quick as for the ordinary two dimensional case. 
More than a hundred significant figures for $z_{SG_3}$ can be obtained when $m$ in Eq. (\ref{zsg3}) is equal to seven.

For the four-dimensional Sierpinski gasket $SG_4(n)$, we use the following definitions.

\bigskip

\begin{defi} \label{defisg4} Consider the four-dimensional Sierpinski gasket $SG_4(n)$ at stage $n$. (a) Define $M_4(n) \equiv N_{DM}(SG_4(n))$ as the number of dimer-monomers. (b) Define $f_4(n)$ as the number of dimer-monomers such that the five outmost vertices are occupied by monomers. (c) Define $g_4(n)$ as the number of dimer-monomers such that one of the outmost vertices is occupied by a dimer and the other four outmost vertices are occupied by monomers. (d) Define $h_4(n)$ as the number of dimer-monomers such that two of the outmost vertices are occupied by dimers and the other three outmost vertices are occupied by monomers. (e) Define $r_4(n)$ as the number of dimer-monomers such that two of the outmost vertices are occupied by monomers and the other three outmost vertices are occupied by dimers. (f) Define $s_4(n)$ as the number of dimer-monomers such that one of the outmost vertices is occupied by a monomer and the other four outmost vertices are occupied by dimers. (g) Define $t_4(n)$ as the number of dimer-monomers such that all five outmost vertices are occupied by dimers.
\end{defi}

\bigskip

\begin{figure}[htbp]
\unitlength 1.8mm 
\begin{picture}(58,9)
\put(2,0){\line(1,0){6}}
\put(2,0){\line(4,3){8}}
\put(2,0){\line(-1,3){2}}
\put(2,0){\line(1,3){3}}
\put(8,0){\line(-1,3){3}}
\put(8,0){\line(1,3){2}}
\put(8,0){\line(-4,3){8}}
\put(0,6){\line(1,0){10}}
\put(5,9){\line(5,-3){5}}
\put(5,9){\line(-5,-3){5}}
\put(5,4.4){\circle{2}}
\put(5,-2){\makebox(0,0){$M_4(n)$}}
\put(18,0){\line(1,0){6}}
\put(18,0){\line(4,3){8}}
\put(18,0){\line(-1,3){2}}
\put(18,0){\line(1,3){3}}
\put(24,0){\line(-1,3){3}}
\put(24,0){\line(1,3){2}}
\put(24,0){\line(-4,3){8}}
\put(16,6){\line(1,0){10}}
\put(21,9){\line(5,-3){5}}
\put(21,9){\line(-5,-3){5}}
\put(18.5,0.5){\circle{1}}
\put(23.5,0.5){\circle{1}}
\put(17,5.7){\circle{1}}
\put(25,5.7){\circle{1}}
\put(21,8){\circle{1}}
\put(21,-2){\makebox(0,0){$f_4(n)$}}
\put(34,0){\line(1,0){6}}
\put(34,0){\line(4,3){8}}
\put(34,0){\line(-1,3){2}}
\put(34,0){\line(1,3){3}}
\put(40,0){\line(-1,3){3}}
\put(40,0){\line(1,3){2}}
\put(40,0){\line(-4,3){8}}
\put(32,6){\line(1,0){10}}
\put(37,9){\line(5,-3){5}}
\put(37,9){\line(-5,-3){5}}
\put(34.5,0.5){\circle{1}}
\put(39.5,0.5){\circle{1}}
\put(33,5.7){\circle{1}}
\put(41,5.7){\circle{1}}
\put(37,8){\circle*{1}}
\put(37,-2){\makebox(0,0){$g_4(n)$}}
\put(50,0){\line(1,0){6}}
\put(50,0){\line(4,3){8}}
\put(50,0){\line(-1,3){2}}
\put(50,0){\line(1,3){3}}
\put(56,0){\line(-1,3){3}}
\put(56,0){\line(1,3){2}}
\put(56,0){\line(-4,3){8}}
\put(48,6){\line(1,0){10}}
\put(53,9){\line(5,-3){5}}
\put(53,9){\line(-5,-3){5}}
\put(50.5,0.5){\circle*{1}}
\put(55.5,0.5){\circle*{1}}
\put(49,5.7){\circle{1}}
\put(57,5.7){\circle{1}}
\put(53,8){\circle{1}}
\put(53,-2){\makebox(0,0){$h_4(n)$}}
\end{picture}

\vspace*{10mm}

\begin{picture}(58,9)
\put(18,0){\line(1,0){6}}
\put(18,0){\line(4,3){8}}
\put(18,0){\line(-1,3){2}}
\put(18,0){\line(1,3){3}}
\put(24,0){\line(-1,3){3}}
\put(24,0){\line(1,3){2}}
\put(24,0){\line(-4,3){8}}
\put(16,6){\line(1,0){10}}
\put(21,9){\line(5,-3){5}}
\put(21,9){\line(-5,-3){5}}
\put(18.5,0.5){\circle*{1}}
\put(23.5,0.5){\circle*{1}}
\put(17,5.7){\circle{1}}
\put(25,5.7){\circle{1}}
\put(21,8){\circle*{1}}
\put(21,-2){\makebox(0,0){$r_4(n)$}}
\put(34,0){\line(1,0){6}}
\put(34,0){\line(4,3){8}}
\put(34,0){\line(-1,3){2}}
\put(34,0){\line(1,3){3}}
\put(40,0){\line(-1,3){3}}
\put(40,0){\line(1,3){2}}
\put(40,0){\line(-4,3){8}}
\put(32,6){\line(1,0){10}}
\put(37,9){\line(5,-3){5}}
\put(37,9){\line(-5,-3){5}}
\put(34.5,0.5){\circle*{1}}
\put(39.5,0.5){\circle*{1}}
\put(33,5.7){\circle*{1}}
\put(41,5.7){\circle*{1}}
\put(37,8){\circle{1}}
\put(37,-2){\makebox(0,0){$s_4(n)$}}
\put(50,0){\line(1,0){6}}
\put(50,0){\line(4,3){8}}
\put(50,0){\line(-1,3){2}}
\put(50,0){\line(1,3){3}}
\put(56,0){\line(-1,3){3}}
\put(56,0){\line(1,3){2}}
\put(56,0){\line(-4,3){8}}
\put(48,6){\line(1,0){10}}
\put(53,9){\line(5,-3){5}}
\put(53,9){\line(-5,-3){5}}
\put(50.5,0.5){\circle*{1}}
\put(55.5,0.5){\circle*{1}}
\put(49,5.7){\circle*{1}}
\put(57,5.7){\circle*{1}}
\put(53,8){\circle*{1}}
\put(53,-2){\makebox(0,0){$t_4(n)$}}
\end{picture}

\vspace*{5mm}
\caption{\footnotesize{Illustration for the spanning subgraphs $M_4(n)$, $f_4(n)$, $g_4(n)$, $h_4(n)$, $r_4(n)$, $s_4(n)$ and $t_4(n)$. Only the five outmost vertices are shown explicitly for $f_4(n)$, $g_4(n)$, $h_4(n)$, $r_4(n)$, $s_4(n)$ and $t_4(n)$, where each open circle is occupied by a monomer and each solid circle is occupied by a dimer.}} 
\label{fghrstfig}
\end{figure}
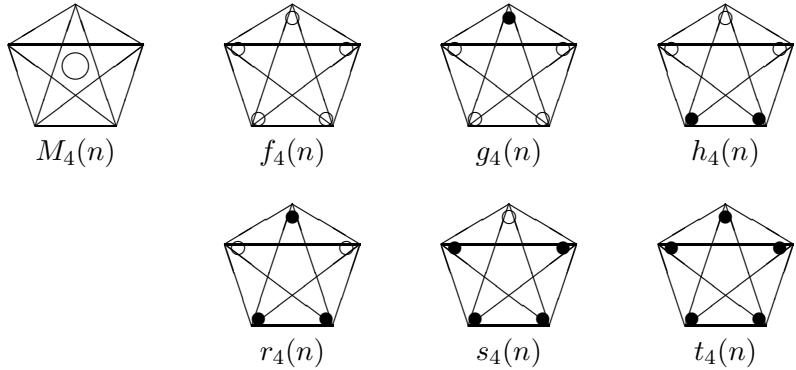

The quantities $M_4(n)$, $f_4(n)$, $g_4(n)$, $h_4(n)$, $r_4(n)$, $s_4(n)$ and $t_4(n)$ are illustrated in Fig. \ref{fghrstfig}, where only the outmost vertices are shown. There are ${5 \choose 1}=5$ equivalent $g_4(n)$, ${5 \choose 2}=10$ equivalent $h_4(n)$, ${5 \choose 3}=10$ equivalent $r_4(n)$, ${5 \choose 1}=5$ equivalent $s_4(n)$. By definition,
\beqs
M_4(n) & = & f_4(n)+5g_4(n)+10h_4(n)+10r_4(n)+5s_4(n)+t_4(n) \ . \cr & &
\label{fsg4}
\eeqs
The initial values at stage zero are $f_4(0)=1$, $g_4(0)=0$, $h_4(0)=1$, $r_4(0)=0$, $s_4(0)=3$, $t_4(0)=0$ and $M_4(0)=26$.

\bigskip

\bigskip

We write a computer program to obtain the recursion relations for $SG_4(n)$. They are too lengthy to be included here, and are available from the authors on request. Some values of $M_4(n)$, $f_4(n)$, $g_4(n)$, $h_4(n)$, $r_4(n)$, $s_4(n)$, $t_4(n)$ are listed in Table \ref{tablesg4}. These numbers grow exponentially, and do not have simple integer factorizations.

\bigskip

\begin{table}[htbp]
\caption{\label{tablesg4} The first few values of $M_4(n)$, $f_4(n)$, $g_4(n)$, $h_4(n)$, $r_4(n)$, $s_4(n)$, $t_4(n)$.}
\begin{center}
\begin{tabular}{|c||r|r|r|}
\hline\hline 
$n$      &  0 &       1 & 2 \\ \hline \hline
$M_4(n)$ & 26 & 141,339 & 1,567,220,397,434,550,336,692,928 \\ \hline 
$f_4(n)$ &  1 &   2,460 &    27,951,923,701,499,685,610,752 \\ \hline 
$g_4(n)$ &  0 &   3,168 &    34,593,006,758,221,606,500,864 \\ \hline 
$h_4(n)$ &  1 &   3,990 &    42,806,033,106,111,666,338,688 \\ \hline
$r_4(n)$ &  0 &   4,852 &    52,961,649,817,161,203,920,896 \\ \hline 
$s_4(n)$ &  3 &   5,683 &    65,517,552,720,775,495,239,744 \\ \hline 
$t_4(n)$ &  0 &   6,204 &    81,038,847,105,336,439,783,296 \\ 
\hline\hline 
\end{tabular}
\end{center}
\end{table}

\bigskip

Define $\alpha_4(n)=g_4(n)/f_4(n)$ and $\gamma_4(n)=t_4(n)/s_4(n)$ as in Eq. (\ref{ratiodef}). 
We find $\{\alpha_4(n)\}_{n=1}^\infty$ decreases monotonically and $\{\gamma_4(n)\}_{n=1}^\infty$ increases monotonically with $1 \le \gamma_4(n) \le \alpha_4(n)$ for positive integer $n$.
The values of $\alpha_4(n)$, $\gamma_4(n)$ and other ratios for small $n$ are listed in Table \ref{tablesg4n}. 

\bigskip

\begin{table}[htbp]
\caption{\label{tablesg4n} The first few values of $\alpha_4(n)$, $\gamma_4(n)$ and other ratios. The last digits given are rounded off.}
\begin{center}
\begin{tabular}{|c||r|r|r|r|}
\hline\hline 
$n$              &                1 &                2 &                3 & 4 \\ \hline\hline 
$\alpha_4(n)$    & 1.28780487804878 & 1.23758948141253 & 1.23734576161280 & 1.23734575732423 \\ \hline 
$h_4(n)/g_4(n)$  & 1.25946969696970 & 1.23741869000555 & 1.23734575860203 & 1.23734575732423 \\ \hline 
$r_4(n)/h_4(n)$  & 1.21604010025063 & 1.23724732179398 & 1.23734575559125 & 1.23734575732423 \\ \hline 
$s_4(n)/r_4(n)$  & 1.17126957955482 & 1.23707537334960 & 1.23734575258048 & 1.23734575732423 \\ \hline 
$\gamma_4(n)$    & 1.09167693119831 & 1.23690284114717 & 1.23734574956971 & 1.23734575732423 \\ \hline\hline 
\end{tabular}
\end{center}
\end{table}

\bigskip

By a similar argument as Lemma \ref{lemmasg2b}, the asymptotic growth constant for the number of dimer-monomers on $SG_4(n)$ is bounded:
\beq
\frac{2\ln f_4(m) +5\ln [1+2\gamma_4(m)]}{5^{m+1}}  \le z_{SG_4} \le \frac{2\ln f_4(m) +5\ln [1+2\alpha_4(m)]}{5^{m+1}}   \ ,
\label{zsg4}
\eeq
with $m$ a positive integer. We have the following proposition.

\bigskip

\begin{propo} \label{proposg4} The asymptotic growth constant for the number of dimer-monomers on the four-dimensional Sierpinski gasket $SG_4(n)$ in the large $n$ limit is $z_{SG_4}=0.876779402949...$.

\end{propo}

\bigskip

The convergence of the upper and lower bounds is as quick as for the ordinary two dimensional case. 
More than a hundred significant figures for $z_{SG_4}$ can be obtained when $m$ in Eq. (\ref{zsg4}) is equal to seven.

\section{Summary}
\label{sectionVI}

The bounds of the asymptotic growth constants for dimer-monomers on $SG_2(n)$, $SG_3(n)$ and $SG_4(n)$ given in sections \ref{sectionIII} and \ref{sectionV} lead to the following conjecture for general $SG_d(n)$.

\bigskip

\begin{conj} \label{conjsgd} Define $\alpha_d(n)$ as the ratio: the number of dimer-monomers on $SG_d(n)$ with all but one outmost vertices covered by monomers divided by that with all outmost vertices covered by monomers; define $\gamma_d(n)$ as the ratio: the number of dimer-monomers on $SG_d(n)$ with all outmost vertices covered by dimers divided by that with all but one outmost vertices covered by dimers.
The asymptotic growth constant for the number of dimer-monomers on the $d$-dimensional Sierpinski gasket $SG_d$ is bounded
\beq
\frac{2\ln f_d(m) +(d+1)\ln [1+2\gamma_d(m)]}{(d+1)^{m+1}} \le z_{SG_d} \le \frac{2\ln f_d(m) +(d+1)\ln [1+2\alpha_d(m)]}{(d+1)^{m+1}}   \ .
\label{zsgd}
\eeq
\end{conj}

\bigskip

We notice that the convergence of the upper and lower bounds of the asymptotic growth constants for dimer-monomers on $SG_d(n)$ is about the same for each integer $d \ge 2$, in contrast to the results observed in \cite{sfs} for spanning forests on $SG_d(n)$ where the convergence of the bounds of the asymptotic growth constants becomes slow when $d$ increases.

The values of $z_{SG_d}$ increases as dimension $d$ increases. Similarly
for the generalized two-dimensional Sierpinski gasket, the values of $z_{SG_{2,b}}$ increases slightly as $b$ increases.

Compare the present results with those in Ref. \cite{sts}, we find that the number of dimer-monomers on the Sierpinski gasket $SG_d(n)$ is less than that of spanning trees in general.

\appendix

\section{Recursion relations for $SG_{2,3}(n)$}

We give the recursion relations for the generalized two-dimensional Sierpinski gasket $SG_{2,3}(n)$ here. For any non-negative integer $n$, we have
\beqs
\lefteqn{f_{2,3}(n+1)} \cr\cr
& = & f_{2,3}^6(n) + 15f_{2,3}^5(n)g_{2,3}(n) + 12f_{2,3}^5(n)h_{2,3}(n) + 84f_{2,3}^4(n)g_{2,3}^2(n) + 3f_{2,3}^5(n)t_{2,3}(n) \cr\cr & & + 117f_{2,3}^4(n)g_{2,3}(n)h_{2,3}(n) + 220f_{2,3}^3(n)g_{2,3}^3(n) + 24f_{2,3}^4(n)g_{2,3}(n)t_{2,3}(n) + 33f_{2,3}^4(n)h_{2,3}^2(n) \cr\cr & & + 390f_{2,3}^3(n)g_{2,3}^2(n)h_{2,3}(n) + 273f_{2,3}^2(n)g_{2,3}^4(n) + 9f_{2,3}^4(n)h_{2,3}(n)t_{2,3}(n) \cr\cr & & + 63f_{2,3}^3(n)g_{2,3}^2(n)t_{2,3}(n) + 180f_{2,3}^3(n)g_{2,3}(n)h_{2,3}^2(n) + 519f_{2,3}^2(n)g_{2,3}^3(n)h_{2,3}(n) \cr\cr & & + 141f_{2,3}(n)g_{2,3}^5(n) + 36f_{2,3}^3(n)g_{2,3}(n)h_{2,3}(n)t_{2,3}(n) + 60f_{2,3}^2(n)g_{2,3}^3(n)t_{2,3}(n) \cr\cr & & + 20f_{2,3}^3(n)h_{2,3}^3(n) + 264f_{2,3}^2(n)g_{2,3}^2(n)h_{2,3}^2(n) + 240f_{2,3}(n)g_{2,3}^4(n)h_{2,3}(n) + 20g_{2,3}^6(n) \cr\cr & & + 3f_{2,3}^3(n)h_{2,3}^2(n)t_{2,3}(n) + 30f_{2,3}^2(n)g_{2,3}^2(n)h_{2,3}(n)t_{2,3}(n) + 15f_{2,3}(n)g_{2,3}^4(n)t_{2,3}(n) \cr\cr & & + 33f_{2,3}^2(n)g_{2,3}(n)h_{2,3}^3(n) + 90f_{2,3}(n)g_{2,3}^3(n)h_{2,3}^2(n) + 21g_{2,3}^5(n)h_{2,3}(n) \ ,
\label{f23eq}
\eeqs
\beqs
\lefteqn{g_{2,3}(n+1)} \cr\cr
& = & f_{2,3}^5(n)g_{2,3}(n) + 2f_{2,3}^5(n)h_{2,3}(n) + 13f_{2,3}^4(n)g_{2,3}^2(n) + f_{2,3}^5(n)t_{2,3}(n) \cr\cr & & + 35f_{2,3}^4(n)g_{2,3}(n)h_{2,3}(n) + 60f_{2,3}^3(n)g_{2,3}^3(n) + 14f_{2,3}^4(n)g_{2,3}(n)t_{2,3}(n) + 18f_{2,3}^4(n)h_{2,3}^2(n) \cr\cr & & + 188f_{2,3}^3(n)g_{2,3}^2(n)h_{2,3}(n) + 120f_{2,3}^2(n)g_{2,3}^4(n) + 11f_{2,3}^4(n)h_{2,3}(n)t_{2,3}(n) \cr\cr & & + 61f_{2,3}^3(n)g_{2,3}^2(n)t_{2,3}(n) + 152f_{2,3}^3(n)g_{2,3}(n)h_{2,3}^2(n) + 397f_{2,3}^2(n)g_{2,3}^3(n)h_{2,3}(n) \cr\cr & & + 99f_{2,3}(n)g_{2,3}^5(n) + f_{2,3}^4(n)t_{2,3}^2(n) + 72f_{2,3}^3(n)g_{2,3}(n)h_{2,3}(n)t_{2,3}(n) \cr\cr & & + 102f_{2,3}^2(n)g_{2,3}^3(n)t_{2,3}(n) + 30f_{2,3}^3(n)h_{2,3}^3(n) + 372f_{2,3}^2(n)g_{2,3}^2(n)h_{2,3}^2(n) \cr\cr & & + 310f_{2,3}(n)g_{2,3}^4(n)h_{2,3}(n) + 25g_{2,3}^6(n) + 4f_{2,3}^3(n)g_{2,3}(n)t_{2,3}^2(n) + 13f_{2,3}^3(n)h_{2,3}^2(n)t_{2,3}(n) \cr\cr & & + 130f_{2,3}^2(n)g_{2,3}^2(n)h_{2,3}(n)t_{2,3}(n) + 57f_{2,3}(n)g_{2,3}^4(n)t_{2,3}(n) + 107f_{2,3}^2(n)g_{2,3}(n)h_{2,3}^3(n) \cr\cr & & + 266f_{2,3}(n)g_{2,3}^3(n)h_{2,3}^2(n) + 63g_{2,3}^5(n)h_{2,3}(n) + 4f_{2,3}^2(n)g_{2,3}^2(n)t_{2,3}^2(n) \cr\cr & & + 30f_{2,3}^2(n)g_{2,3}(n)h_{2,3}^2(n)t_{2,3}(n) + 52f_{2,3}(n)g_{2,3}^3(n)h_{2,3}(n)t_{2,3}(n) + 6g_{2,3}^5(n)t_{2,3}(n) \cr\cr & & + 6f_{2,3}^2(n)h_{2,3}^4(n) + 60f_{2,3}(n)g_{2,3}^2(n)h_{2,3}^3(n) + 34g_{2,3}^4(n)h_{2,3}^2(n) \ , 
\label{g23eq}
\eeqs
\beqs
\lefteqn{h_{2,3}(n+1)} \cr\cr 
& = & f_{2,3}^4(n)g_{2,3}^2(n) + 4f_{2,3}^4(n)g_{2,3}(n)h_{2,3}(n) + 11f_{2,3}^3(n)g_{2,3}^3(n) + 2f_{2,3}^4(n)g_{2,3}(n)t_{2,3}(n) \cr\cr & & + 4f_{2,3}^4(n)h_{2,3}^2(n) + 50f_{2,3}^3(n)g_{2,3}^2(n)h_{2,3}(n) + 40f_{2,3}^2(n)g_{2,3}^4(n) + 4f_{2,3}^4(n)h_{2,3}(n)t_{2,3}(n) \cr\cr & & + 21f_{2,3}^3(n)g_{2,3}^2(n)t_{2,3}(n) + 68f_{2,3}^3(n)g_{2,3}(n)h_{2,3}^2(n) + 191f_{2,3}^2(n)g_{2,3}^3(n)h_{2,3}(n) \cr\cr & & + 56f_{2,3}(n)g_{2,3}^5(n) + f_{2,3}^4(n)t_{2,3}^2(n) + 52f_{2,3}^3(n)g_{2,3}(n)h_{2,3}(n)t_{2,3}(n) \cr\cr & & + 66f_{2,3}^2(n)g_{2,3}^3(n)t_{2,3}(n) + 25f_{2,3}^3(n)h_{2,3}^3(n) + 289f_{2,3}^2(n)g_{2,3}^2(n)h_{2,3}^2(n) \cr\cr & & + 263f_{2,3}(n)g_{2,3}^4(n)h_{2,3}(n) + 24g_{2,3}^6(n) + 9f_{2,3}^3(n)g_{2,3}(n)t_{2,3}^2(n) + 23f_{2,3}^3(n)h_{2,3}^2(n)t_{2,3}(n) \cr\cr & & + 167f_{2,3}^2(n)g_{2,3}^2(n)h_{2,3}(n)t_{2,3}(n) + 72f_{2,3}(n)g_{2,3}^4(n)t_{2,3}(n) + 150f_{2,3}^2(n)g_{2,3}(n)h_{2,3}^3(n) \cr\cr & & + 390f_{2,3}(n)g_{2,3}^3(n)h_{2,3}^2(n) + 101g_{2,3}^5(n)h_{2,3}(n) + 5f_{2,3}^3(n)h_{2,3}(n)t_{2,3}^2(n) \cr\cr & & + 19f_{2,3}^2(n)g_{2,3}^2(n)t_{2,3}^2(n) + 94f_{2,3}^2(n)g_{2,3}(n)h_{2,3}^2(n)t_{2,3}(n) + 158f_{2,3}(n)g_{2,3}^3(n)h_{2,3}(n)t_{2,3}(n) \cr\cr & & + 20g_{2,3}^5(n)t_{2,3}(n) + 20f_{2,3}^2(n)h_{2,3}^4(n) + 201f_{2,3}(n)g_{2,3}^2(n)h_{2,3}^3(n) + 123g_{2,3}^4(n)h_{2,3}^2(n) \cr\cr & & + 11f_{2,3}^2(n)g_{2,3}(n)h_{2,3}(n)t_{2,3}^2(n) + 9f_{2,3}(n)g_{2,3}^3(n)t_{2,3}^2(n) + 9f_{2,3}^2(n)h_{2,3}^3(n)t_{2,3}(n) \cr\cr & & + 68f_{2,3}(n)g_{2,3}^2(n)h_{2,3}^2(n)t_{2,3}(n) + 27g_{2,3}^4(n)h_{2,3}(n)t_{2,3}(n) + 27f_{2,3}(n)g_{2,3}(n)h_{2,3}^4(n) \cr\cr & & + 41g_{2,3}^3(n)h_{2,3}^3(n) \ ,
\label{h23eq}
\eeqs
\beqs
\lefteqn{t_{2,3}(n+1)} \cr\cr
& = & f_{2,3}^3(n)g_{2,3}^3(n) + 6f_{2,3}^3(n)g_{2,3}^2(n)h_{2,3}(n) + 9f_{2,3}^2(n)g_{2,3}^4(n) + 3f_{2,3}^3(n)g_{2,3}^2(n)t_{2,3}(n) \cr\cr & & + 12f_{2,3}^3(n)g_{2,3}(n)h_{2,3}^2(n) + 57f_{2,3}^2(n)g_{2,3}^3(n)h_{2,3}(n) + 24f_{2,3}(n)g_{2,3}^5(n) \cr\cr & & + 12f_{2,3}^3(n)g_{2,3}(n)h_{2,3}(n)t_{2,3}(n) + 24f_{2,3}^2(n)g_{2,3}^3(n)t_{2,3}(n) + 8f_{2,3}^3(n)h_{2,3}^3(n) \cr\cr & & + 126f_{2,3}^2(n)g_{2,3}^2(n)h_{2,3}^2(n) + 150f_{2,3}(n)g_{2,3}^4(n)h_{2,3}(n) + 20g_{2,3}^6(n) + 3f_{2,3}^3(n)g_{2,3}(n)t_{2,3}^2(n) \cr\cr & & + 12f_{2,3}^3(n)h_{2,3}^2(n)t_{2,3}(n) + 99f_{2,3}^2(n)g_{2,3}^2(n)h_{2,3}(n)t_{2,3}(n) + 51f_{2,3}(n)g_{2,3}^4(n)t_{2,3}(n) \cr\cr & & + 111f_{2,3}^2(n)g_{2,3}(n)h_{2,3}^3(n) + 324f_{2,3}(n)g_{2,3}^3(n)h_{2,3}^2(n) + 120g_{2,3}^5(n)h_{2,3}(n) \cr\cr & & + 6f_{2,3}^3(n)h_{2,3}(n)t_{2,3}^2(n) + 18f_{2,3}^2(n)g_{2,3}^2(n)t_{2,3}^2(n) + 117f_{2,3}^2(n)g_{2,3}(n)h_{2,3}^2(n)t_{2,3}(n) \cr\cr & & + 186f_{2,3}(n)g_{2,3}^3(n)h_{2,3}(n)t_{2,3}(n) + 33g_{2,3}^5(n)t_{2,3}(n) + 30f_{2,3}^2(n)h_{2,3}^4(n) \cr\cr & & + 282f_{2,3}(n)g_{2,3}^2(n)h_{2,3}^3(n) + 240g_{2,3}^4(n)h_{2,3}^2(n) + f_{2,3}^3(n)t_{2,3}^3(n) \cr\cr & & + 36f_{2,3}^2(n)g_{2,3}(n)h_{2,3}(n)t_{2,3}^2(n) + 21f_{2,3}(n)g_{2,3}^3(n)t_{2,3}^2(n) + 33f_{2,3}^2(n)h_{2,3}^3(n)t_{2,3}(n) \cr\cr & & + 183f_{2,3}(n)g_{2,3}^2(n)h_{2,3}^2(n)t_{2,3}(n) + 99g_{2,3}^4(n)h_{2,3}(n)t_{2,3}(n) + 87f_{2,3}(n)g_{2,3}(n)h_{2,3}^4(n) \cr\cr & & + 180g_{2,3}^3(n)h_{2,3}^3(n) + 3f_{2,3}^2(n)g_{2,3}(n)t_{2,3}^3(n) + 9f_{2,3}^2(n)h_{2,3}^2(n)t_{2,3}^2(n) \cr\cr & & + 24f_{2,3}(n)g_{2,3}^2(n)h_{2,3}(n)t_{2,3}^2(n) + 6g_{2,3}^4(n)t_{2,3}^2(n) + 42f_{2,3}(n)g_{2,3}(n)h_{2,3}^3(n)t_{2,3}(n) \cr\cr & & + 63g_{2,3}^3(n)h_{2,3}^2(n)t_{2,3}(n) + 6f_{2,3}(n)h_{2,3}^5(n) + 39g_{2,3}^2(n)h_{2,3}^4(n) \ . 
\label{t23eq}
\eeqs
There are always $2916=4\times3^6$ terms in these equations.

\section{Recursion relations for $SG_3(n)$}

We give the recursion relations for the three-dimensional Sierpinski gasket $SG_3(n)$ here. For any non-negative integer $n$, we have
\beqs
\lefteqn{f_3(n+1)} \cr\cr
& = & f_3^4(n) + 12f_3^3(n)g_3(n) + 12f_3^3(n)h_3(n) + 48f_3^2(n)g_3^2(n) + 4f_3^3(n)r_3(n) \cr\cr
& & + 84f_3^2(n)g_3(n)h_3(n) + 72f_3(n)g_3^3(n) + 24f_3^2(n)g_3(n)r_3(n) + 30f_3^2(n)h_3^2(n) \cr\cr
& & + 156f_3(n)g_3^2(n)h_3(n) + 30g_3^4(n) + 12f_3^2(n)h_3(n)r_3(n) + 36f_3(n)g_3^2(n)r_3(n) \cr\cr
& & + 84f_3(n)g_3(n)h_3^2(n) + 60g_3^3(n)h_3(n) + 24f_3(n)g_3(n)h_3(n)r_3(n) + 8g_3^3(n)r_3(n) \cr\cr
& & + 8f_3(n)h_3^3(n) + 24g_3^2(n)h_3^2(n) \ , 
\label{f3eq}
\eeqs
\beqs
\lefteqn{g_3(n+1)} \cr\cr
& = & f_3^3(n)g_3(n) + 3f_3^3(n)h_3(n) + 9f_3^2(n)g_3^2(n) + 3f_3^3(n)r_3(n) + 33f_3^2(n)g_3(n)h_3(n) \cr\cr
& & + 24f_3(n)g_3^3(n) + f_3^3(n)s_3(n) + 24f_3^2(n)g_3(n)r_3(n) + 21f_3^2(n)h_3^2(n) + 96f_3(n)g_3^2(n)h_3(n) \cr\cr
& & + 18g_3^4(n) + 6f_3^2(n)g_3(n)s_3(n) + 21f_3^2(n)h_3(n)r_3(n) + 51f_3(n)g_3^2(n)r_3(n) \cr\cr
& & + 93f_3(n)g_3(n)h_3^2(n) + 69g_3^3(n)h_3(n) + 3f_3^2(n)h_3(n)s_3(n) + 9f_3(n)g_3^2(n)s_3(n) \cr\cr
& & + 3f_3^2(n)r_3^2(n) + 66f_3(n)g_3(n)h_3(n)r_3(n) + 24g_3^3(n)r_3(n) + 21f_3(n)h_3^3(n) + 66g_3^2(n)h_3^2(n) \cr\cr
& & + 6f_3(n)g_3(n)h_3(n)s_3(n) + 2g_3^3(n)s_3(n) + 6f_3(n)g_3(n)r_3^2(n) + 12f_3(n)h_3^2(n)r_3(n) \cr\cr
& & + 24g_3^2(n)h_3(n)r_3(n) + 14g_3(n)h_3^3(n) \ ,
\label{g3eq}
\eeqs
\beqs
\lefteqn{h_3(n+1)} \cr\cr
& = & f_3^2(n)g_3^2(n) + 6f_3^2(n)g_3(n)h_3(n) + 6f_3(n)g_3^3(n) + 6f_3^2(n)g_3(n)r_3(n) + 8f_3^2(n)h_3^2(n) \cr\cr
& & + 38f_3(n)g_3^2(n)h_3(n) + 8g_3^4(n) + 2f_3^2(n)g_3(n)s_3(n) + 14f_3^2(n)h_3(n)r_3(n) \cr\cr
& & + 30f_3(n)g_3^2(n)r_3(n) + 64f_3(n)g_3(n)h_3^2(n) + 50g_3^3(n)h_3(n) + 4f_3^2(n)h_3(n)s_3(n) \cr\cr
& & + 8f_3(n)g_3^2(n)s_3(n) + 5f_3^2(n)r_3^2(n) + 80f_3(n)g_3(n)h_3(n)r_3(n) + 30g_3^3(n)r_3(n) \cr\cr
& & + 26f_3(n)h_3^3(n) + 87g_3^2(n)h_3^2(n) + 2f_3^2(n)r_3(n)s_3(n) + 16f_3(n)g_3(n)h_3(n)s_3(n) \cr\cr
& & + 6g_3^3(n)s_3(n) + 18f_3(n)g_3(n)r_3^2(n) + 34f_3(n)h_3^2(n)r_3(n) + 72g_3^2(n)h_3(n)r_3(n) \cr\cr
& & + 44g_3(n)h_3^3(n) + 4f_3(n)g_3(n)r_3(n)s_3(n) + 4f_3(n)h_3^2(n)s_3(n) + 8g_3^2(n)h_3(n)s_3(n) \cr\cr
& & + 8f_3(n)h_3(n)r_3^2(n) + 8g_3^2(n)r_3^2(n) + 28g_3(n)h_3^2(n)r_3(n) + 4h_3^4(n) \ , 
\label{h3eq}
\eeqs
\beqs
\lefteqn{r_3(n+1)} \cr\cr
& = & f_3(n)g_3^3(n) + 9f_3(n)g_3^2(n)h_3(n) + 3g_3^4(n) + 9f_3(n)g_3^2(n)r_3(n) + 24f_3(n)g_3(n)h_3^2(n) \cr\cr
& & + 27g_3^3(n)h_3(n) + 3f_3(n)g_3^2(n)s_3(n) + 42f_3(n)g_3(n)h_3(n)r_3(n) + 22g_3^3(n)r_3(n) \cr\cr
& & + 18f_3(n)h_3^3(n) + 75g_3^2(n)h_3^2(n) + 12f_3(n)g_3(n)h_3(n)s_3(n) + 6g_3^3(n)s_3(n) \cr\cr
& & + 15f_3(n)g_3(n)r_3^2(n) + 39f_3(n)h_3^2(n)r_3(n) + 99g_3^2(n)h_3(n)r_3(n) + 69g_3(n)h_3^3(n) \cr\cr
& & + 6f_3(n)g_3(n)r_3(n)s_3(n) + 9f_3(n)h_3^2(n)s_3(n) + 21g_3^2(n)h_3(n)s_3(n) + 21f_3(n)h_3(n)r_3^2(n) \cr\cr
& & + 24g_3^2(n)r_3^2(n) + 96g_3(n)h_3^2(n)r_3(n) + 15h_3^4(n) + 6f_3(n)h_3(n)r_3(n)s_3(n) \cr\cr
& & + 6g_3^2(n)r_3(n)s_3(n) + 12g_3(n)h_3^2(n)s_3(n) + 2f_3(n)r_3^3(n) + 24g_3(n)h_3(n)r_3^2(n) \cr\cr
& & + 14h_3^3(n)r_3(n) \ , 
\label{r3eq}
\eeqs
\beqs
\lefteqn{s_3(n+1)} \cr\cr
& = & g_3^4(n) + 12g_3^3(n)h_3(n) + 12g_3^3(n)r_3(n) + 48g_3^2(n)h_3^2(n) + 4g_3^3(n)s_3(n) \cr\cr
& & + 84g_3^2(n)h_3(n)r_3(n) + 72g_3(n)h_3^3(n) + 24g_3^2(n)h_3(n)s_3(n) + 30g_3^2(n)r_3^2(n) \cr\cr
& & + 156g_3(n)h_3^2(n)r_3(n) + 30h_3^4(n) + 12g_3^2(n)r_3(n)s_3(n) + 36g_3(n)h_3^2(n)s_3(n) \cr\cr
& & + 84g_3(n)h_3(n)r_3^2(n) + 60h_3^3(n)r_3(n) + 24g_3(n)h_3(n)r_3(n)s_3(n) + 8h_3^3(n)s_3(n) \cr\cr
& & + 8g_3(n)r_3^3(n) + 24h_3^2(n)r_3^2(n) \ .
\label{s3eq}
\eeqs
There are always $729=3^6$ terms in these equations.

\begin{acknowledgments}
The authors would like to thank Weigen Yan for helpful discussion. The research of S.C.C. was partially supported by the NSC grant NSC-95-2112-M-006-004. The research of L.C.C was partially supported by the NSC grant NSC-95-2115-M-030-002.
\end{acknowledgments}


\begin{thebibliography}{00}

\bibitem{gaunt69} 
D. S. Gaunt, {\it Phys. Rev.} {\bf 179}, 174 (1969).

\bibitem{Heilmann70} 
O. J. Heilmann and E. H. Lieb, {\it Phys. Rev. Lett.} {\bf 24}, 1412 (1970).

\bibitem{Heilmann72} 
O. J. Heilmann and E. H. Lieb, {\it Commun. Math. Phys.} {\bf 25}, 190 (1972).

\bibitem{kasteleyn61} 
P. W. Kasteleyn, {\it Physica (Amsterdam)} {\bf 27}, 1209 (1961).

\bibitem{temperley61} 
H. N. V. Temperley and M. E. Fisher, {\it Philos. Mag.} {\bf 6}, 1061 (1961).

\bibitem{fisher61} 
M. E. Fisher, {\it Phys. Rev.} {\bf 124}, 1664 (1961); {\bf 132}, 1411 (1963).

\bibitem{jerrum} 
M. Jerrum, {\it J. Stat. Phys.} {\bf 48}, 121 (1987); {\bf 59}, 1087 (1990).

\bibitem{lu99} 
W. T. Lu and F. Y. Wu, {\it Phys. Lett. A} {\bf 259}, 108 (1999).

\bibitem{tzeng03} 
W.-J. Tzeng and F. Y. Wu, {\it J. Stat. Phys.} {\bf 110}, 671 (2003).

\bibitem{izmailian03} 
N. Sh. Izmailian, K. B. Oganesyan and C. K. Hu, {\it Phys. Rev. E} {\bf 67}, 066114 (2003).

\bibitem{izmailian05} 
N. Sh. Izmailian, V. B. Priezzhev, P. Ruelle and C. K. Hu, {\it Phys. Rev. Lett.} {\bf 95}, 260602 (2005).

\bibitem{yan05} 
W. G. Yan, Y.-N. Yeh and F. J. Zhang, {\it Int. J. Quantum Chem.} {\bf 105}, 124 (2005).

\bibitem{yan06} 
W. G. Yan and Y.-N. Yeh, {\it Science in China A: Math.} {\bf 49}, 1383 (2006).

\bibitem{kong06} 
Y. Kong, {\it Phys. Rev. E} {\bf 73}, 016106 (2006).

\bibitem{izmailian06} 
N. Sh. Izmailian, K. B. Oganesyan, M.-C. Wu and C. K. Hu, {\it Phys. Rev. E} {\bf 73}, 016128 (2006).

\bibitem{wu06} 
F. Y. Wu, {\it Phys. Rev. E} {\bf 74}, 020104(R) (2006); {\bf 74}, 039907(E) (2006).

\bibitem{kong06n} 
Y. Kong, {\it Phys. Rev. E} {\bf 74}, 011102 (2006).

\bibitem{kong06nn} 
Y. Kong, {\it Phys. Rev. E} {\bf 74}, 061102 (2006)

\bibitem{mandelbrot}
B. B. Mandelbrot, {\it The Fractal Geometry of Nature}, Freeman, San Francisco, 1982.

\bibitem{Falconer}
K. J. Falconer, {\it Fractal Geometry: Mathematical Foundations and Applications}, 2nd ed., Wiley, Chichester, 2003.

\bibitem{Gefen80}
Y. Gefen, B. B. Mandelbrot and A. Aharony, {\it Phys. Rev. Lett.} {\bf 45}, 855 (1980).

\bibitem{Gefen81}
Y. Gefen, A. Aharony, B. B. Mandelbrot and S. Kirkpartrick, {\it Phys. Rev. Lett.} {\bf 47}, 1771 (1981).

\bibitem{Rammal}
R. Rammal and G. Toulouse, {\it Phys. Rev. Lett.} {\bf 49}, 1194 (1982).

\bibitem{Alexander}
S. Alexander, {\it Phys. Rev. B} {\bf 27}, 1541 (1983).

\bibitem{Domany}
E. Domany, S. Alexander, D. Bensimon and L. P. Kadanoff, {\it Phys. Rev. B} {\bf 28}, 3110 (1983).

\bibitem{Gefen8384}
Y. Gefen, A. Aharony and B. B. Mandelbrot, {\it J. Phys. A: Math. Gen.} {\bf 16}, 1267 (1983); Y. Gefen, A. Aharony, Y. Shapir and B. B. Mandelbrot, {\it ibid.} {\bf 17}, 435 (1984); Y. Gefen, A. Aharony and B. B. Mandelbrot, {\it ibid.} {\bf 17}, 1277 (1984).

\bibitem{Guyer}
R. A. Guyer, {\it Phys. Rev. A} {\bf 29}, 2751 (1984).

\bibitem{Kusuoka}
K. Hattori, T. Hattori and S. Kusuoka, {\it Probab. Theory Relat. Fields} {\bf 84}, 1 (1990); T. Hattori and S. Kusuoka, {\it ibid.} {\bf 93}, 273 (1992).

\bibitem{Dhar97}
D. Dhar and A. Dhar, {\it Phys. Rev. E} {\bf 55}, R2093 (1997).

\bibitem{Daerden}
F. Daerden and C. Vanderzande, {\it Physica A} {\bf 256}, 533 (1998).

\bibitem{Dhar05}
D. Dhar, {\it Phys. Rev. E} {\bf 71}, 031801 (2005).

\bibitem{bbook}
N. L. Biggs, {\it Algebraic Graph Theory}, 2nd ed., Cambridge University Press, Cambridge, 1993.

\bibitem{fh} 
F. Harary, {\it Graph Theory}, Addison-Wesley, New York, 1969.

\bibitem{Hilfer}
R. Hilfer and A. Blumen, {\it J. Phys. A: Math. Gen.} {\bf 17}, L537 (1984).

\bibitem{sfs}
S.-C. Chang and L.-C. Chen, math-ph/0612083.

\bibitem{sts}
S.-C. Chang and L.-C. Chen, {\it J. Stat. Phys.}, in press.

\end{thebibliography}
\end{document}